\documentclass[12pt]{spieman}  
\usepackage{amsmath,amsfonts,amssymb}
\usepackage{graphicx}
\usepackage{setspace}
\usepackage{tocloft}
\usepackage{lineno}
\usepackage[acronym,shortcuts]{glossaries}
\setacronymstyle{long-short}
\usepackage{subcaption}
\usepackage{tabularx}
\usepackage{multirow}

\newacronym{pat}{PAT}{photoacoustic tomography}
\newacronym{pa}{PA}{photoacoustic}
\newacronym{hsi}{HSI}{hyperspectral imaging}
\newacronym{hs}{HS}{hyperspectral}
\newacronym{hb}{Hb}{deoxyhaemoglobin}
\newacronym{hbo2}{HbO$_2$}{oxyhaemoglobin}
\newacronym{oxy}{sO$_2$}{oxygen saturation}
\newacronym{lsu}{LSU}{linear spectral unmixing}
\newacronym{dis}{DIS}{double-integrating sphere}
\newacronym{iad}{IAD}{inverse adding-doubling}
\newacronym{mua}{$\mu_a$}{absorption coefficient}
\newacronym{mus}{$\mu_s$}{scattering coefficient}
\newacronym{nnls}{NNLS}{non-negative least squares}
\newacronym{ood}{OOD}{out-of-distribution}
\newacronym{mae}{MAE}{mean absolute error}
\newacronym{roi}{ROI}{region of interest}
\newacronym{r}{R-value}{Pearson correlation coefficient}
\newacronym{mri}{MRI}{magnetic resonance imaging}

\title{Anthropomorphic tissue-mimicking phantoms for oximetry validation in multispectral optical imaging}

\author[a,b,*]{Kris K. Dreher}
\author[c,d]{Janek Gröhl}
\author[a,e]{Friso Grace}
\author[a,f]{Leonardo Ayala}
\author[a,g]{Jan-Hinrich Nölke}
\author[a,g]{Christoph J. Bender}
\author[c,d]{Melissa J. Watt}
\author[c,d]{Catherine-Louise White}
\author[c,d]{Ran Tao}
\author[i,j]{Wibke Johnen}
\author[a]{Minu D. Tizabi}
\author[a]{Alexander Seitel}
\author[a,f,g,h,$\dagger$,*]{Lena Maier-Hein}
\author[c,d,$\dagger$,*]{Sarah E. Bohndiek}
\affil[a]{Division of Intelligent Medical Systems (IMSY), German Cancer Research Center (DKFZ), Heidelberg, Germany}
\affil[b]{Faculty of Physics and Astronomy, Heidelberg University, Heidelberg, Germany}
\affil[c]{University of Cambridge, CRUK Cambridge Institute, Cambridge, United Kingdom}
\affil[d]{University of Cambridge, Department of Physics, Cambridge, United Kingdom}
\affil[e]{School of Physics and Astronomy, University of St Andrews, St Andrews, United Kingdom}
\affil[f]{National Center for Tumor Diseases (NCT), NCT Heidelberg, a partnership between DKFZ and University Hospital Heidelberg, Heidelberg, Germany}
\affil[g]{Faculty of Mathematics and Computer Science, Heidelberg University, Heidelberg, Germany}
\affil[h]{Medical Faculty, Heidelberg University, Heidelberg, Germany}
\affil[i]{Division of Medical Physics in Radiation Oncology, DKFZ, Heidelberg, Germany}
\affil[j]{National Center for Radiation Research in Oncology (NCRO), Heidelberg Institute for Radiation Oncology (HIRO), Heidelberg, Germany}

\cftpagenumbersoff{figure}
\cftpagenumbersoff{table} 
\begin{document} 
\maketitle

\begin{abstract}

\textbf{Significance:} Optical imaging of blood oxygenation (sO$_2$) can be achieved based on the differential absorption spectra of oxy- and deoxy-haemoglobin. A key challenge in realising clinical validation of the sO$_2$ biomarkers is the absence of reliable sO$_2$ reference standards, including test objects.

\noindent \textbf{Aim:} To enable quantitative testing of multispectral imaging methods for assessment of sO$_2$ by introducing anthropomorphic phantoms with appropriate tissue-mimicking optical properties.

\noindent \textbf{Approach:} We used the stable copolymer-in-oil base material to create physical anthropomorphic structures and optimised dyes to mimic the optical absorption of blood across a wide spectral range. Using  3D-printed phantom moulds generated from a magnetic resonance image of a human forearm, we moulded the material into an anthropomorphic shape. Using both reflectance hyperspectral imaging (HSI) and photoacoustic tomography (PAT), we acquired images of the forearm phantoms and evaluated the performance of linear spectral unmixing (LSU).

\noindent \textbf{Results:} Based on 10 fabricated forearm phantoms with vessel-like structures featuring five distinct sO$_2$ levels (between 0 and 100\%), we showed that the measured absorption spectra of the material correlated well with HSI and PAT data with a Pearson correlation coefficient consistently above 0.8. Further, the application of LSU enabled a quantification of the mean absolute error in sO$_2$ assessment with HSI and PAT.

\noindent \textbf{Conclusion:} Our anthropomorphic tissue-mimicking phantoms hold potential to provide a robust tool for developing, standardising, and validating optical imaging of sO$_2$. 
 
\end{abstract}


{\noindent \footnotesize\textbf{*}Corresponding authors: KKD, LMH \{k.dreher, l.maier-hein\}@dkfz-heidelberg.de, SEB \linkable{seb53@cam.ac.uk} }\newline
{\noindent \footnotesize\textbf{$\dagger$}Shared last authorship}

\begin{spacing}{2}   

\section{Introduction}
\label{sect:intro}  
\par
\Ac*{hbo2} and \ac*{hb} are critical endogenous contrast agents that enable noninvasive measurement of blood \ac*{oxy}, also referred to as oximetry, due to their distinct optical absorption spectra. Oximetry methods are invaluable for a range of clinical applications throughout the patient care pathway, from diagnosis to treatment planning, and monitoring of treatment response \cite{taylor-williams_noninvasive_2022,lefebvre_potential_2022,riksen_photoacoustic_2023,ntziachristos_addressing_2024,park_clinical_2024}. Pulse oximetry is the most widely available of these tools, which employs red and infrared light to noninvasively estimate \ac*{oxy} for bulk tissue at a single measurement site, such as a fingertip\cite{taylor-williams_noninvasive_2022}. 
\par
To obtain spatially resolved information, for example in image-guided surgery, imaging modalities like \ac*{hsi} and \ac*{pat} are used. \ac*{hsi} and \ac*{pat} typically involve making wavelength-resolved measurements across the visible and near-infrared spectral range \cite{taylor-williams_noninvasive_2022}. \Ac*{hsi} uses light diffusely reflected from the tissue to map \ac*{hbo2} and \ac*{hb} signals near the tissue surface\cite{lu_medical_2014,clancy_surgical_2020}, whereas \ac*{pat} combines pulsed laser illumination and ultrasonic detection to probe deeper tissue layers\cite{wang_biomedical_2007,wang_photoacoustic_2017}. In both techniques, deriving \ac*{oxy} from spectral data commonly relies on \ac*{lsu}\cite{hochuli_estimating_2019,bioucas-dias_hyperspectral_2012}, assuming that the optical absorption responsible for the image contrast is a linear combination of the absorption spectra of all contrast agents present at any given point weighted by their concentration.

Accurately quantifying \ac*{oxy} in \ac*{hsi} and \ac*{pat} is challenging, since a range of assumptions are made in both the datacube (x,y,$\lambda$) reconstruction and spectral analysis pipeline that can lead to corruption of the measured tissue biomarkers. For example, in \ac*{hsi}, signals can be distorted due to additional optical interactions, such as fluorescence, while in \ac*{pat}, depth-dependent signal attenuation arises, known as spectral colouring. In both modalities, patient motion and skin tone bias can introduce further complexity\cite{else_effects_2023,hochuli_estimating_2019}. 
Consequently, developing robust oximetry calibration methods for \ac*{hsi} and \ac*{pat} remains an active area of research\cite{ayala_band_2022,larsson_artificial_2024,grohl_learned_2021,nolke_photoacoustic_2024}.
\par
For the development and rigorous validation of any scientific method, including oximetry, a reliable performance measure or reference is essential. In oximetry, however, the principal challenge is that an \textit{in vivo} ground truth for \ac*{oxy} is not available noninvasively with current technology.
Therefore, many studies only rely on qualitative visualisations or measurement of relative changes in the same individual or specimen over time, rather than calibrating for absolute \ac*{oxy}\cite{chen_simultaneous_2019,ayala_live_2019,kirchner_photoacoustics_2019}.
Validation approaches can take several forms. Validation can be developed by comparison to simulated data, however, these often fail to generalise when applied \textit{in vivo} tissue\cite{grohl_deep_2021}. Adding a level of complexity, experimental data obtained from bulk test objects ('phantoms') made with a mixture of blood, haemoglobin or other biological solutions\cite{zalev_opto-acoustic_2019,kirchner_multiple_2021} can closely mimic tissue spectra and facilitate validation, but typically: do not allow well-controlled adjustment of \ac*{oxy}, lack internal structure, tend to be unstable over time, and are prone to bacterial contamination\cite{hacker_criteria_2022,walter_development_2023}. Addressing some of these limitations, blood flow phantoms allow for chemical manipulation of \ac*{oxy} as blood flows through a circuit while being imaged \cite{vogt_photoacoustic_2019,ghassemi_rapid_2015}. Nevertheless, blood flow circuits rely on ancillary reference measurements (e.g., partial pressure probes or offline oximeters) to provide a gold standard reference and are typically simple in their structure e.g. a tube flowing through a slab of base material.
\par
Anthropomorphic phantoms with stable, tissue-mimicking optical properties over a broad wavelength range remain rare \cite{dantuma_tunable_2021}. Obtaining oximetry estimations from anthropomorphic phantoms typically requires embedding a flow circuit within a complex material composition, which is prohibitively challenging. To overcome this limitation and tackle the challenge of oximetry validation in \ac*{hsi} and \ac*{pat}, we introduce an anthropomorphic phantom design that mimics the morphology of a human forearm and includes absorbing dyes that replicate the \ac*{mua} characteristics of \ac*{hbo2} and \ac*{hb} between 700 nm and 850 nm. We first select and optimise dye proxies for \ac*{hb} and \ac*{hbo2}. We then created 3D-printable moulds derived from human forearm imaging to achieve morphologically relevant geometry, which are available open-source. Using a copolymer-in-oil base material, we created structurally stable anthropomorphic forearm phantoms with optical characteristics confirmed via simulation studies and signal correlation analyses. Finally, we demonstrated the value of the created phantoms for oximetry validation in \ac*{hsi} and \ac*{pat}. These phantoms provide a versatile, morphologically realistic test platform for oximetry validation in the near-infrared.

\section{Materials and methods}

\begin{figure}[!ht]
    \centering
    \includegraphics[width=0.8\textwidth]{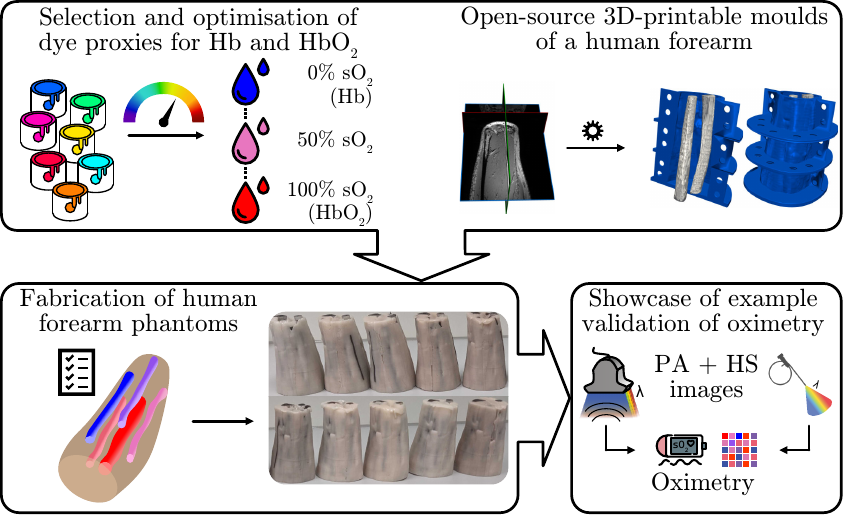}
    \caption{\textbf{Workflow and key contributions of this study.} (Top) First, 26 dyes were investigated to mimic the absorption spectra of \acrfull*{hbo2} and \acrfull*{hb} in the wavelength ($\lambda$) range of 700\,nm to 850\,nm. After a selection and optimisation process, two proxy dyes that could be mixed to 5 levels of \acrfull*{oxy} (0\%, 30\%, 50\%, 70\%, 100\%) were found. Second, for realistic tissue morphology, a 3D-printable mould was created based on an open-source magnetic resonance (MR) image of a human forearm. (bottom) Third, ten forearm phantoms were fabricated and finally, \acrfull*{pa} and \acrfull*{hs} images were acquired and we show that example images of these can be used to validate oximetry methods.
    }
    \label{fig:concept}
\end{figure}

Combining dye proxies with anthropomorphic phantom designs in this study enables a new approach to oximetry validation (Fig. \ref{fig:concept}).

\subsection{Forearm phantom fabrication}
\label{sec:phantom}
\subsubsection{Phantom material}

The base phantom material was prepared according to protocols outlined by Hacker et al.\cite{hacker_copolymer--oil_2021} and Gröhl et al. \cite{grohl_moving_2024}. Briefly, for each $\sim$80\,mL batch of base material, 76.5\,mg of titanium dioxide (TiO$_2$, Sigma Aldrich 232033-100g) was sonicated in a water bath together with 50\,mL of mineral oil (Sigma Aldrich 330779-1L) until completely dispersed. Next, 12.57\,g of polystyrene-block-poly(ethylene-ran-butylene)-block-polystyrene (SEBS, Sigma Aldrich 200557-250G) and 1\,g of butylated hydroxytoluene (HT, Sigma Aldrich W218405-1KG-K) were added to the oil. The mixture was heated in a silicone oil bath at 160°C for about 45 minutes, stirred every 10 minutes, and allowed to liquefy fully. Finally, the beaker was placed in a vacuum chamber to remove any residual air bubbles. A more detailed manufacturing protocol is provided in the (Supplementary Notes S1).

\subsubsection{Optical property characterisation}

The optical properties of the phantom materials were quantified using a \ac*{dis} system (according to the method of Pickering et al. \cite{pickering_double-integrating-sphere_1993} with the system described in Hacker et al. \cite{hacker_copolymer--oil_2021}). The \ac*{dis} system measured total reflectance and transmittance over a wavelength range of 700 to 850\,nm, and the resulting data were processed with the \ac*{iad}\cite{prahl_everything_2011} algorithm to obtain \ac*{mua} and \ac*{mus}\cite{jacques_optical_2013}. The refractive index was set to n=1.4, and the anisotropy factor to g=0.7, as suggested by Jones and Munro\cite{jones_stability_2018}. For each material batch, two optical sample slabs were fabricated. The thickness of each slab was measured five times at three different locations (top, middle, bottom) using a digital calliper and provided as input to the \ac*{iad} algorithm. Subsequently, each location was measured from both the front and back with the \ac*{dis} system, producing 12 individual data points per wavelength.

\subsubsection{Proxy dyes for oxy- and deoxyhemoglobin}

A total of 26 candidate dyes were evaluated for their optical properties in the above-mentioned base material between 700 and 850\,nm (Supplementary Figures S1-3, Table S1). The \ac*{mua} spectra measured with the \ac*{dis} system served as inputs to a \ac*{nnls} optimisation, which aimed to identify a linear combination of dyes whose summed absorption profiles closely match those of \ac*{hbo2} and \ac*{hb}. Based on these results, the most promising dyes were selected and further refined by iteratively adjusting concentrations and re-measuring \ac*{mua}. 

\subsubsection{3D Forearm model}

\begin{figure}[!ht]
    \centering
    \includegraphics[width=\textwidth]{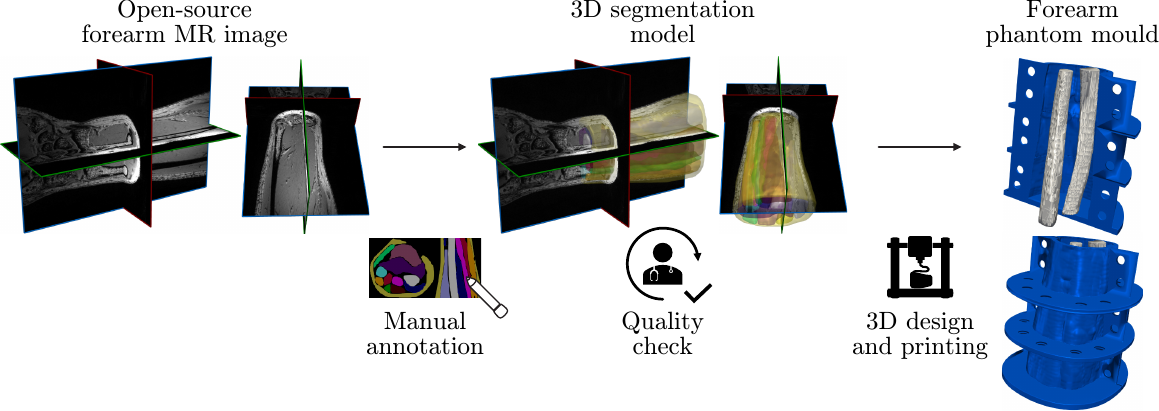}
    \caption{\textbf{3D-printable models were constructed from an open-source magnetic resonance (MR) image.} From left to right: The MR image was manually segmented for bones, muscles and fat. A physician performed a quality check ensuring morphological correctness. Based on this segmentation, a 3D-printable model of the outer hull of the forearm has been designed and printed including the two bones radius and ulna that will stay in the mould as positives.
    }
    \label{fig:mould}
\end{figure}

A high-resolution \ac*{mri} dataset of the human forearm was obtained from Kerkhof et al.\cite{kerkhof_digital_2018}. From these data, muscle and bone structures were segmented, and the segmentation was verified by a physician (MDT). The outer hull and bones were then 3D-printed to serve as mould and embedded features, respectively (Fig. \ref{fig:mould}). The digital model for the mould was designed and constructed using Autodesk Inventor 2024 and Geomagic Freeform. An Objet 500 Connex (Stratasys, Ltd., Eden Prairie, Minnesota, USA) was used for 3D printing. Specifically, VeroCyan$^{\text{TM}}$ was used to fabricate the external mould, while VeroClear$^{\text{TM}}$ was chosen to print the bones, which remained inside each phantom as positive structures. 

\subsubsection{Phantom fabrication process}

Ten forearm phantoms were fabricated, each containing background material and 14 embedded vessels. The vessels were created by drawing phantom material into tubes with diameters of 3, 4, 5, or 6\,mm, producing 140 vessels in total (28 per \ac*{oxy} level, 7 per diameter). At least one vessel for each \ac*{oxy} level was positioned near the surface (“superficial vessel”) in each phantom. Dye mixtures within the vessels were determined based on the outcome of the \ac*{nnls} optimisation. Particular attention was paid to achieving an isosbestic point near 800\,nm. For mixing intermediate \ac*{oxy} levels, the base dyes were accurately weighed, and the resulting ratios were verified by measuring \ac*{mua} spectra and applying \ac*{lsu}. This two-step approach ensured reliable verification of the intended mixing ratios.
\par
Nine of the phantoms were manufactured with background materials spanning three \ac*{oxy} levels (0\%, 50\%, 100\%) and their volumes as fractions of the respective material batch (1\%, 2.5\%, 4\%), yielding 3×3 combinations. An additional \ac*{ood} phantom was created with three distinct combinations of volume fraction and \ac*{oxy} (0.5\%, 100\%), (5\%, 0\%), and (3\%, 70\%) (Supplementary Table S2). Optical properties were verified for each background material (Supplementary Figures S4-7). The fabrication proceeded by splitting the 3D-printed mould into two halves and pouring two 100\,mL batches of the phantom mixture in layers. Vessels were placed incrementally within each half after each layer, and their approximate positions were documented. Once both halves were filled and all 14 vessels were in place, the two parts of the mould were joined, and any remaining cavity in the centre was filled with the residual phantom material.

\subsection{Optical imaging techniques}
\label{sec:imaging}

\subsubsection{Photoacoustic tomography}

All phantoms were scanned using the MSOT Acuity Echo system (iThera Medical GmbH, Munich, Germany) in a water bath using the wavelengths from 700\,nm - 850\,nm in steps of 10\,nm. Each of the ten phantoms was imaged at three predefined locations, and for each location, eight angular views were acquired in 45° increments around the phantom. This arrangement resulted in 24 images per phantom, for a total of 240 images overall. The phantoms were mounted on a rotational stage within the water bath to facilitate consistent data acquisition across all angles and locations.

The acquired time-series data were corrected for laser energy (Supplementary Fig. S8) and filtered using a bandpass with cutoffs at 50 kHz and 20 MHz. The 700\,nm wavelength was excluded from subsequent analysis due to laser instability. Reconstructions were performed with a delay-and-sum algorithm implemented in the open-source toolkit for simulation and image processing for photonics and acoustics (SIMPA)\cite{grohl_simpa_2022}, specifying a speed of sound of 1497.4\,ms$^{-1}$ and a voxel resolution of 0.1\,mm. After reconstruction, a Hilbert transform was applied for envelope detection. The speed of sound was chosen to enable co-registration with concurrently acquired ultrasound images. 

\subsubsection{Hyperspectral imaging}

All phantoms were imaged using the Tivita 2.0 camera (Diaspective Vision GmbH, Am Salzhaff, Germany). The wavelengths from 500\,nm - 1000\,nm in steps of 1\,nm were imaged but for consistency with PAT, we used the wavelengths from 700\,nm - 850\,nm in steps of 10\,nm for analysis in this work. Each phantom was placed on a rotational stage and scanned at eight angular positions in 45° increments. Before each capture, the camera was refocused to ensure clear images. This procedure yielded 8 \ac*{hs} images per phantom, for a total of 80 images across all phantoms. The acquired \ac*{hs} images were automatically corrected for white and dark references using the open-source htc\cite{sellner_hyperspectral_2025} software. 

\subsection{Quality assurance}
\label{sec:qa}

To ensure that our phantoms faithfully replicate the intended optical and acoustic properties, we performed two key evaluations: simulation studies to investigate the effects of air bubbles and speed of sound variations on \ac*{pa} images and signal correlation analyses to confirm the relationship between measured signals and the known absorption spectra at varying \ac*{oxy} levels. 

\subsubsection{Simulation studies}

Ultrasound segmentations including observed air bubbles were used to simulate phantom images, comparing vessel spectra with and without air inclusions. Additionally, a sensitivity analysis on speed of sound variations ($\pm$50 and $\pm$100\,ms$^{-1}$ around the assumed speed of sound 1470 ms$^{-1}$) was performed. Since the main focus of this work is on the optical properties, refer to Supplementary section S3 for detailed experiment descriptions and results. 
Briefly, all simulations were conducted with SIMPA\cite{grohl_simpa_2022}, employing MCX\cite{fang_monte_2009} for photon transport and k-Wave\cite{treeby_k-wave_2010} for acoustic wave propagation. Each simulation used a digital device twin of the MSOT Acuity Echo, and a digital tissue twin constructed from manual segmentations of five representative phantom PA images. The vessels and background regions in these digital twins were assigned the absorption and scattering coefficients obtained via the \ac*{dis} system. 

\subsubsection{Signal correlation}

\Ac*{pa} signals ($S$) are proportional to the product of the Grüneisen parameter ($\Gamma$), \ac*{mua}, and the local fluence ($\Phi$): $S \propto \Gamma \mu_a \Phi$\cite{wang_biomedical_2007}.
Because the Grüneisen parameter of the mixed dyes could be different such that the \ac*{pat} signal does not correlate linearly with the \ac*{mua}\cite{cox_quantitative_2012}, we investigated whether the measured \ac*{mua} for different \ac*{oxy} levels correlates linearly with the \ac*{pa} signal. The processing steps included:

\begin{itemize}
    \item Vessel segmentation: Superficial vessels were segmented as \acp*{roi}, and the top 5\% of brightest pixels in each vessel region were averaged.
    \item Spectrum fitting: A linear regression (one multiplicative factor plus one offset) was applied to match the expected absorption spectrum from 710 nm to 850 nm).
    \item Correlation analysis: The \ac*{r} between the measured \ac*{pa} spectrum and the known absorption spectrum was determined for each vessel.

\end{itemize}

The \ac*{hs} signal ($I$) is formed from the diffusely reflected fraction of light that is neither absorbed nor scattered out of the detection path. A common method to approximate \ac*{mua} from reflectance data uses the Lambert-Beer law, where $\mu_a \propto -log(I)$\cite{oshina_beerlambert_2021}. We therefore checked whether the measured \ac*{mua} for different \ac*{oxy} levels correlated with the \ac*{hs} signal.
The processing steps include:

\begin{itemize}
    \item Region selection: \acp*{roi} were chosen within superficial vessels, avoiding specular highlights and vessel edges to minimise cross-talk from the surrounding tissue.
    \item Lambert-Beer approximation: An approximate absorption spectrum was obtained by applying $\hat{\mu}_a = -log(I)$.
    \item Spectrum fitting: A linear regression (one multiplicative factor plus one offset) was applied to match the expected absorption spectrum from 700 nm to 850 nm.
    \item Correlation analysis: The \ac*{r} was computed to assess the agreement between the derived absorption spectrum and the known phantom absorption.
\end{itemize}

\subsection{Oximetry method validation}
\label{sec:oxy}

To quantify the accuracy of our phantom-based oximetry measurements, the derived spectra (both with and without signal correlation) were used as inputs to a \ac*{lsu} algorithm\cite{hochuli_estimating_2019,bioucas-dias_hyperspectral_2012}. The performance of \ac*{lsu} was assessed by calculating the \ac*{mae} in estimated \ac*{oxy} levels.
In the \ac*{pat} experiments, \ac*{lsu} performance was also evaluated as a function of depth. Additionally, for \ac*{pat} data, a fluence compensation step was introduced before \ac*{lsu} in simulation studies, where the reconstructed image was divided by the estimated fluence to account for optical attenuation effects in the \ac*{pa} signal.
Finally, errors were computed at the most granular level — individual tissue-type instances (e.g., each vessel) - to obtain the mean, confidence interval, and standard deviation. These metrics were then successively aggregated following the hierarchical data structure: across vessels, across tissue types, and ultimately across all phantoms.

\section{Results}

\subsection{Two dyes were found to mimic the optical absorption of haemoglobin across the full range of oxygenations}
\label{sec:res:dyes}

Following the dye testing and optimisation process, IR-1061 and Spectrasense-765 were identified as the closest proxies for blood absorption between 700 and 850\,nm (Fig. \ref{fig:res:blood}, derived \ac*{mus} in Supplementary Fig. S9) with Spectrasense-765 closely reproducing key features of \ac*{hb} (minimum at ~730\,nm, local maximum at ~760\,nm). Although IR-1061 exhibited a less pronounced slope than \ac*{hbo2}, the measured absorption still increased monotonically with wavelength. Notably, the isosbestic point for the two dyes was located at $\sim$800\,nm, mirroring that of \ac*{hbo2} and \ac*{hb}.

\begin{figure}[!ht]
    \centering
    \includegraphics[width=0.8\textwidth]{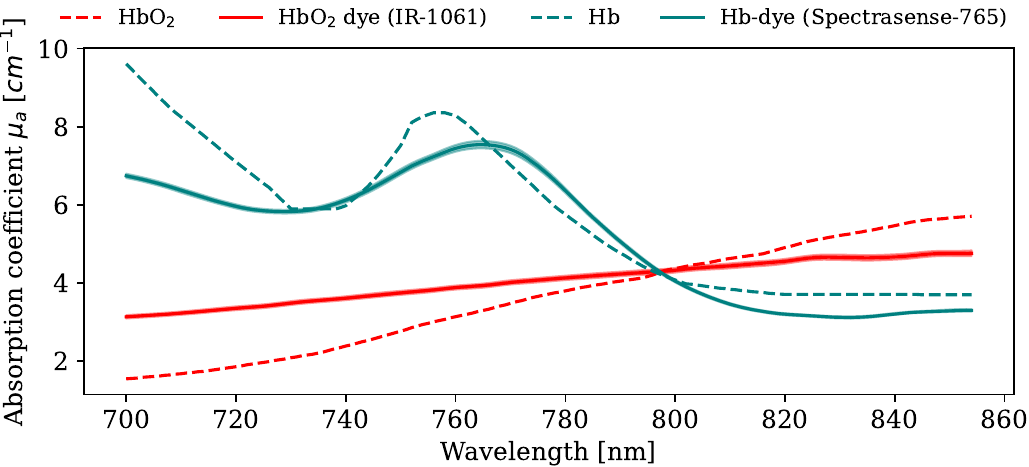}
    \caption{\textbf{IR-1061 and Spectrasense-765 can mimic the absorption characteristics of blood.} Dashed lines indicate the \acrfull*{mua} spectra of \acrfull*{hbo2} and \acrfull*{hb}. Solid lines indicate the measured absorption spectra of the proxy dyes. Bands around the measured spectra indicate the standard deviation across the 12 measurement points for each optical sample slab. Mean absolute errors between targets and proxy dyes are 0.78\,cm$^{-1}$ and 0.64\,cm$^{-1}$ for \ac*{hbo2} and \ac*{hb}, respectively.
    }
    \label{fig:res:blood}
\end{figure}

\subsection{Five oxygen saturation levels with characteristics similar to blood were derived}
\label{sec:res:oxy_levels}
By varying the mixing ratios of IR-1061 and Spectrasense-765, five levels of \ac*{oxy} were created that approximate blood-like characteristics (Fig. \ref{fig:res:oxy_levels}; \ac*{dis} reflectance and transmittance in Supplementary Fig. S10). While the relationship between dye concentrations and resulting \ac*{mua} spectra is highly nonlinear, a relatively linear trend was observed when transitioning from a 90:10 to a 100:0 dye ratio (representing 0\% to 100\% “oxy” levels).  Intermediate mixing ratios (e.g., 93:7, 95:5, and 97:3) then produced five distinct \ac*{oxy} values: 0\%, 30\%, 50\%, 70\%, and 100\%. Taking 90:10 and 100:0 as \ac*{lsu} endmembers yielded intermediate values of 30.7\%, 52.4\%, and 67.4\% \ac*{oxy}.

\begin{figure}[!ht]
    \centering
    \includegraphics[width=\textwidth]{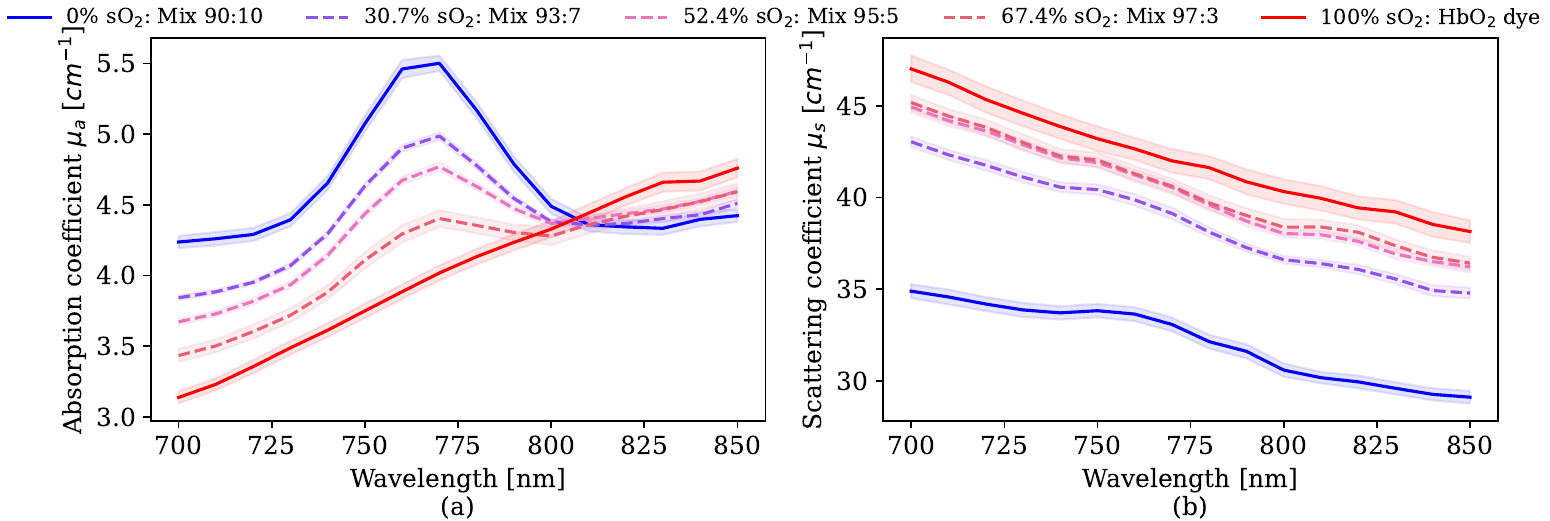}
    \caption{\textbf{Five \acrfull*{oxy} levels were used for forearm phantom fabrication.} Based on IR-1061 and Spectrasense-765, five \ac*{oxy} levels (in \%) were derived with the respective mixture ratios of 100:0, 97:3, 95:5, 93:7, 90:10. (a) and (b) represent the \acrfull*{mua} and \acrfull*{mus}, respectively. Solid lines are the spectra that are used as endmembers (0\% \ac*{oxy} and 100\% \ac*{oxy}) for \acrfull*{lsu}. Dashed lines represent the intermediate levels and the corresponding percentages in the legend are the \ac*{lsu} results when using the solid lines as endmembers. Bands around the spectra indicate the standard deviation across the 12 measurement points for each optical sample slab.
    }
    \label{fig:res:oxy_levels}
\end{figure}

\subsection{Signal correlation shows good agreement with measured absorption spectra}
\label{sec:res:sig_corr}

\begin{figure}[!ht]
    \centering
	\begin{subfigure}[t]{0.49\textwidth}
		\centering
		\includegraphics[width=1\textwidth]{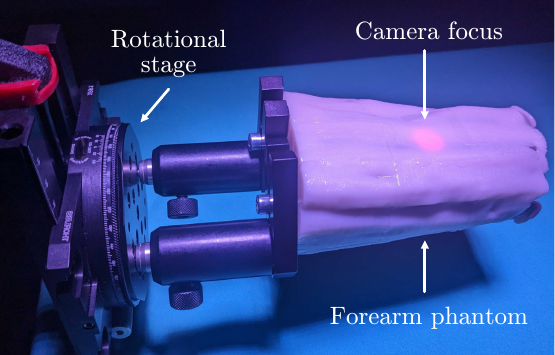}
		\caption{}\label{fig:res:hs_exp:a}		
	\end{subfigure}
	\begin{subfigure}[t]{0.49\textwidth}
		\centering
		\includegraphics[width=0.829\textwidth]{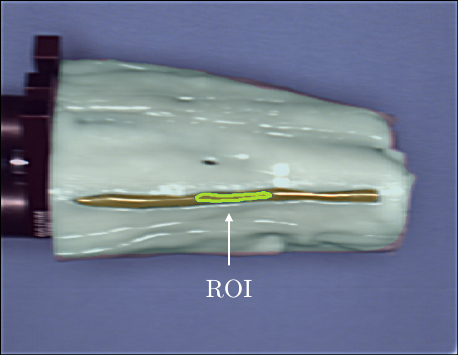}
		\caption{}\label{fig:res:hs_exp:b}
	\end{subfigure}

    \begin{subfigure}[t]{0.8\textwidth}
    \centering
    \includegraphics[width=1\textwidth]{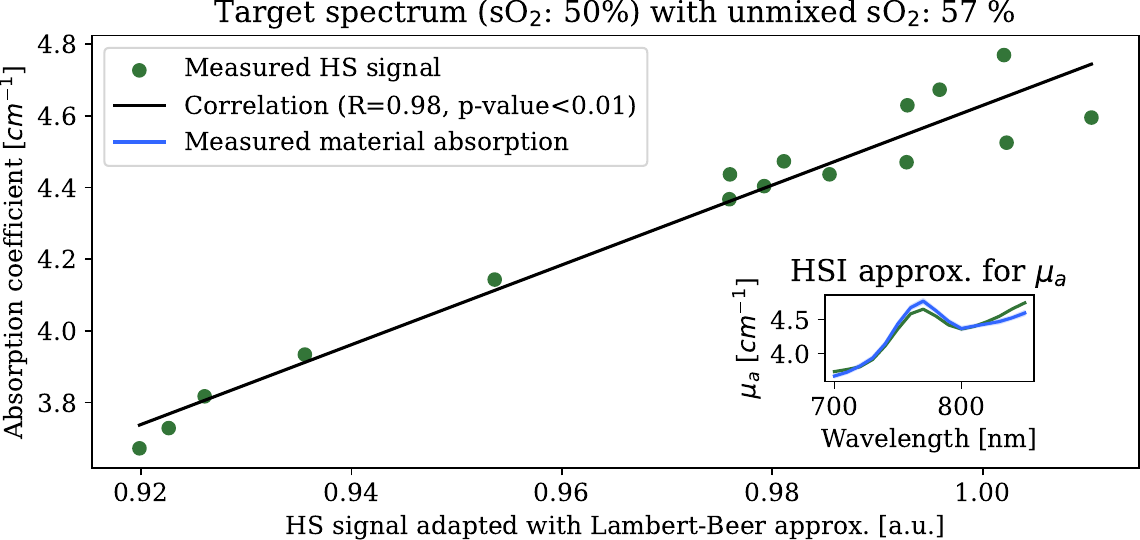}
		\caption{}\label{fig:res:hs_exp:c}		
        
    \end{subfigure}
    \caption{\textbf{Measurement setup for \acrfull*{hsi} (a), corresponding image (b), and signal correlation of the 50\% \acrfull*{oxy} superficial vessel (c)}. The phantoms were mounted on a rotational stage and the camera was adjusted for each phantom such that the middle of the phantoms was in focus (red dot). Images from nine angles in steps of 45° were acquired per phantom. (b) shows an example of an RGB-reconstructed image including a \acrfull*{roi} indicated by yellow margins in the middle of a superficial vessel. The black solid lines in (c) represent the resulting linear regression function with corresponding \acrfull*{r}. The inset plot shows the measured absorption (blue) and estimated absorption (green, using the correlation function) as qualitative confirmation.}
    \label{fig:res:hs_exp}
\end{figure}

The \ac*{hsi} setup (Fig. \ref{fig:res:hs_exp:a}) captures a top-down view of each phantom (Fig. \ref{fig:res:hs_exp:b}). Analysis of the signal in the denoted \ac*{roi} for the representative case with 50\% \ac*{oxy} demonstrated an extremely close match with the expected spectra (Fig. \ref{fig:res:hs_exp:c}).

\begin{figure}[!ht]
    \centering
	\begin{subfigure}[t]{0.35\textwidth}
		\centering
		\includegraphics[width=0.8\textwidth]{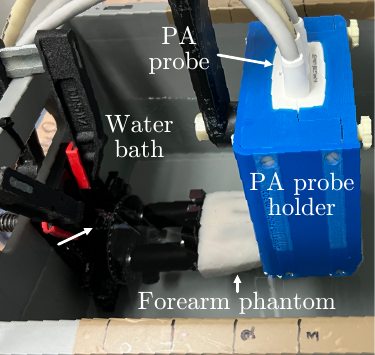}
		\caption{}\label{fig:res:pat_exp:a}		
	\end{subfigure}
	\begin{subfigure}[t]{0.5\textwidth}
		\centering
		\includegraphics[width=\textwidth]{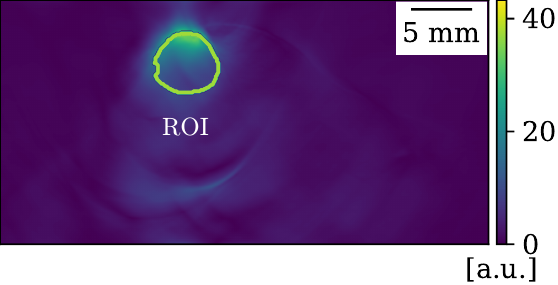}
		\caption{}\label{fig:res:pat_exp:b}
	\end{subfigure}

    \begin{subfigure}[t]{0.8\textwidth}
    \centering
    \includegraphics[width=1\textwidth]{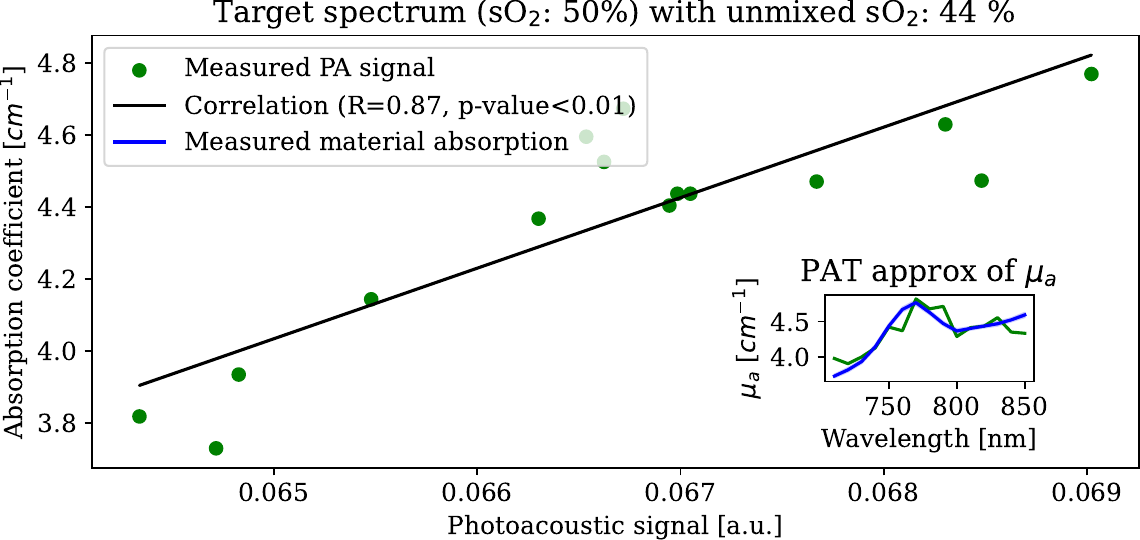}
		\caption{}\label{fig:res:pat_exp:c}		
        
    \end{subfigure}
    
    \caption{\textbf{Measurement setup for \acrfull*{pat} (a), corresponding image (b), and signal correlation of the 50\% \acrfull*{oxy} superficial vessel (c).} The phantoms were mounted on a rotational stage in a water bath and images were acquired from nine angles in steps of 45° in three distinct locations along the phantoms. For each measurement location, the PAT probe which was attached to a mechanical arm to minimise motion during image acquisitions, was adjusted such that it was approximately in the middle of the phantoms and 2\,mm above the uppermost point of the phantom. (b) shows an example PA image including a \acrfull*{roi} indicated by yellow margins in the middle of a superficial vessel. The black solid lines in (c) represent the resulting linear regression function with corresponding \acrfull*{r}. The inset plot shows the measured absorption (blue) and estimated absorption (green, using the correlation function) as qualitative confirmation.}
    \label{fig:res:pat_exp}
\end{figure}

The \ac*{pat} system (Fig. \ref{fig:res:pat_exp:a}) provides cross-sectional images of the phantoms (Fig. \ref{fig:res:pat_exp:b}). The \ac*{pat} correlation analysis in Fig. \ref{fig:res:pat_exp:c} also shows high concordance with the expected absorption, although a higher degree of noise is seen in the \ac*{pat} data compared to \ac*{hsi}. Analogous results were obtained from other imaged forearm phantoms across a range of defined oxygenations (Supplementary Figures S11-S14). The correlation coefficient \ac*{r} between the imaging signals and the measured absorption spectra in superficial vessels across the five example phantoms all exceeded 0.8, indicating strong agreement (Table \ref{tab:corr}). 

\begin{table}
	\begin{center}
		\begin{tabular}{l c c c c c }
			 & 0\% \ac*{oxy} & 30\% \ac*{oxy} & 50\% \ac*{oxy} & 70\% \ac*{oxy} & 100\% \ac*{oxy} \\
			\hline
				HSI & 0.85 & 0.99 & 0.98 & 0.90 & 0.97 \\
                    PAT & 0.85 & 0.83 & 0.87 & 0.93 & 0.99 \\
			\hline
		\end{tabular}
	\end{center}
    \caption{\ac*{r} for linear regression of \acrfull*{hsi} and \acrfull*{pat} signal and absorption correlation.}
    \label{tab:corr}
\end{table}


\subsection{Forearm phantoms can be used for validation of oximetry methods}
\label{sec:res:oxy_val}

\begin{figure}[!ht]
    \centering
    \includegraphics[width=0.8\textwidth]{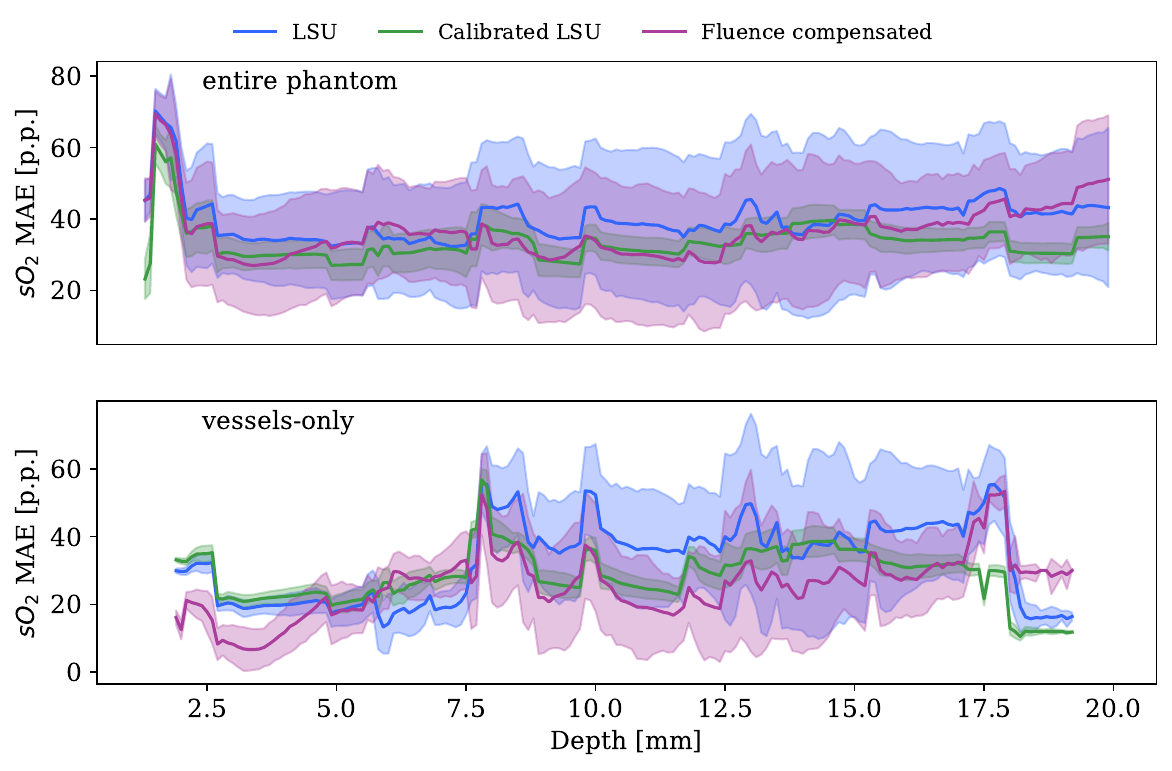}
    \caption{\textbf{Error analysis of three oximetry methods for photoacoustic tomography as a function of depth.} \acrfull*{mae} in percentage points (p.p.) of \acrfull*{lsu} without (blue) and with calibration (green), and with fluence compensation (pink) using a superficial vessel is plotted against the depth of the evaluation pixel for both the whole phantom (top) and for the vessels only (bottom). The bands around the solid lines indicate the standard deviation.
    }
    \label{fig:res:lsu_depth}
\end{figure}

Good performance was found for the three tested oximetry methods (\ac*{lsu}, \ac*{lsu} with calibration from signal correlation analysis and \ac*{lsu} with prior fluence compensation) applied to both \ac*{pat} and \ac*{hsi} data (Table \ref{tab:oxy}). On average, the calibrated versions of \ac*{lsu} showed lower \ac*{mae} and reduced standard deviation compared to the uncalibrated approach, suggesting improved performance. Fluence compensation in \ac*{pat} increased the accuracy of evaluation in vessels, although its benefit for the entire phantom was similar to the uncalibrated method. A key consideration for PAT is evaluation at a function of tissue depths (Fig. \ref{fig:res:lsu_depth}) through which our findings indicate that the calibrated \ac*{lsu} outperforms the uncalibrated approach at most depths in the entire phantom; the fluence compensation method particularly performs well for the first 5\,mm.

\begin{table}
	\begin{center}
		\begin{tabular}{l c c c c }
			& \multicolumn{2}{c}{PAT} & \multicolumn{2}{c}{HSI} \\
			\hline
			& entire phantom & vessels-only & entire phantom & vessels-only \\
                \hline
	 	\multirow{2}{5em}{LSU} & 33.4$\pm$19.6 & 30.4$\pm$18.4 & 39.9$\pm$15.0 & 45.3$\pm$19.0 \\
	 	& [32.6, 34.1] & [29.4, 31.3] & [39.6, 40.3] & [44.7, 46.0] \\
	 	\hline
	 	\multirow{2}{5em}{Calibrated LSU} & 29.9$\pm$5.0 & 27.9$\pm$4.9 & 32.0$\pm$0.6 & 31.1$\pm$0.8 \\
	 	& [29.7, 30.1] & [27.6, 28.1] & [32.0, 32.0] & [31.1, 31.1] \\
	 	\hline
	 	\multirow{2}{5em}{Fluence compensation} & 27.9$\pm$17.9 & 24.2$\pm$16.9 & \multicolumn{2}{c}{\multirow{2}{*}{not applicable}}\\
	 	& [27.2, 28.6] & [23.3, 25.1] & \multicolumn{2}{c}{} \\
                \hline
		\end{tabular}
	\end{center}
    \caption{Mean absolute error for \acrfull*{lsu} applied on \acrfull*{pat} images and \acrfull*{hsi} images. Calibrated \ac*{lsu} indicates that the correlation function was calculated via linear regression in sec. \ref{sec:res:sig_corr}. For \ac*{pat}, Fluence compensation means that the reconstructed image was corrected based on simulated fluence. The standard deviation is indicated with “±” and the bounds of the 95\% confidence intervals of the mean are in square brackets: [low, high]. All values are in percentage points (p.p.).}
    \label{tab:oxy}
\end{table}

\section{Discussion}
\label{sec:disc}
In this study, we presented a set of anthropomorphic forearm phantoms that were designed to replicate the optical absorption of \ac*{hbo2} and \ac*{hb} at five distinct \ac*{oxy} levels (0\%, 30\%, 50\%, 70\%, 100\%) within a biologically relevant wavelength range (700–850\,nm). By employing 3D-printed moulds and strategically selected dyes, these phantoms achieved a morphologically realistic geometry while reproducing the optical properties \ac*{hbo2} and \ac*{hb}. 

Phantom designs of previous work often rely on simplified geometries such as tubular or multilayer structures. While flow-based setups for controlled \ac*{oxy} exist, they generally lack morphological accuracy or long-term stability. Our approach addresses these shortcomings by incorporating forearm morphology and carefully chosen dyes. This strategy yields phantoms that more closely match the morphology and absorption characteristics of human tissue and provides a robust testing platform for oximetry methods across various imaging modalities. It also avoids the need for an \textit{in situ} assessment of ground-truth \ac*{oxy}, as would be required in a flow circuit. While there may be instances when testing with blood itself is vital, for example, for optical systems that rely on scattering from moving red blood cells, for many target applications such as intraoperative imaging, optical systems could be more easily and routinely tested using the approach outlined here.

One core contribution of our work is the open-source data and code for both the phantom moulds and dye optimisation process. Researchers can 3D-print their own moulds derived from high-resolution MRI data, ensuring consistent morphology across laboratories. They can also adapt our dye optimisation framework to derive new chromophore combinations, extending the utility of our work to other tissue types. Moreover, our validation experiments with \ac*{hsi} and \ac*{pat} indicate that the fabricated phantoms produce signals in good agreement with measured absorption, thereby offering a reliable, standardised environment not only for validation of different imaging systems between sites, but also for evaluating different oximetry algorithms. By providing a standardised but anthropomorphic phantom, the approach described here could accelerate the development and comparison of optical imaging methods, ultimately improving clinical translation.

Our correlation analyses show that both \ac*{hsi} and \ac*{pat} signals strongly match the known absorption coefficients of each phantom, underscoring the fidelity of the phantoms for experimental validation. In particular, Pearson correlation coefficients consistently exceeded 0.8, and the analysis for the five example forearms demonstrated close alignment with the expected spectral behaviour. Some instability was observed when applying oximetry methods at depth, which is likely due to various vessels being embedded at different positions, having substantially larger absorption coefficients than their surroundings. Therefore, when going from background material to vessel material, the error of the oximetry method might experience a sudden drop or increase.

Despite these advantages, several limitations remain. First, the two dyes used exhibit highly nonlinear mixing behaviour, complicating the process of achieving intermediate \ac*{oxy} levels. Although we successfully generated five distinct levels, further refinement is needed to enhance reproducibility and minimise iteration. Further, we did not test the long-term stability of these dyes within the base copolymer-in-oil matrix. Second, the \ac*{dis} system employed for optical characterisation is susceptible to measurement uncertainties, which have been reported in prior studies\cite{tao_tutorial_2024}. While we attempted to minimise these uncertainties by verifying the mixing ratio both by weighing the dyes and by using \ac*{lsu} on the measured \ac*{mua}, minor deviations are still possible. Since the focus of this paper was on the optical properties of the phantoms, a full acoustic characterisation was not performed, and only basic speed of sound measurements, borrowed from previous work\cite{hacker_copolymer--oil_2021}, were used. Finally, the phantoms contained small air bubbles despite extensive vacuuming, which may cause acoustic reverberations, particularly for vessels located deeper than 1.5\,cm. Therefore, we recommend limiting the validation of oximetry methods with these phantoms to vessels whose centres lie at depths shallower than 1.5\,cm.

\section{Conclusion}
By enabling accurate and repeatable performance assessments, our tissue-mimicking phantoms provide a robust standard for oximetry validation, bridging the gap left by limited \textit{in vivo} reference methods. Looking ahead, the fabrication strategies and dye selection processes described here offer an initial step for developing even more complex phantoms, thereby fostering more reliable and clinically translatable optical imaging techniques.

\subsection*{Disclosures}
The authors declare no conflict of interest regarding this work.

\subsection* {Code and Data Availability}
\label{sec:code}
The data and code to reproduce the findings of this study are openly available. The data are available under the CC-BY 4.0 license at: \url{https://doi.org/10.5281/zenodo.15102333}. The code is available under the MIT license at: 
\newline
\url{https://github.com/IMSY-DKFZ/anthropomorphic-phantoms}.

\subsection* {Acknowledgments}
This project has received funding from: the European Research Council (ERC) under the European Union’s Horizon 2020 research and innovation programme (grant agreement No. [101002198]); Cancer Research UK (under grant numbers C9545/A29580, C14478/A27855); the UKRI Engineering and Physical Sciences Research Council (EP/L015889/1, EP/X037770/1, EP/V027069/1). RT acknowledges the financial support of the Trinity Barlow Scholarship, Cambridge Trust International Scholarship and the Herchel Smith Scholarship. JG acknowledges funding from the Walter Benjamin Stipendium of the Deutsche Forschungsgemeinschaft. For the purpose of open access, the author has applied a Creative Commons Attribution (CC-BY) licence to any Author Accepted Manuscript version arising. The authors would like to thank Prof. Dr. Jürgen Hesser for the helpful and insightful discussions. A Large Language Model (ChatGPT) has been used to enhance the language of the manuscript.
All model outputs have been thoroughly reviewed and verified for correctness by the authors.


\bibliography{article}   
\bibliographystyle{spiejour}   

\end{spacing}
\end{document}


\maketitle

{\noindent \footnotesize\textbf{*}Corresponding authors: KKD, LMH \{k.dreher,l.maier-hein\}@dkfz-heidelberg.de, SEB \linkable{seb53@cam.ac.uk} }\newline
{\noindent \footnotesize\textbf{$\dagger$}Shared last authorship}

\begin{spacing}{2}   


\section{Supplementary Notes}

The phantom manufacturing protocol used for quality assurance during phantom production is embedded below.

\includegraphics[width=0.93\textwidth,page=1]{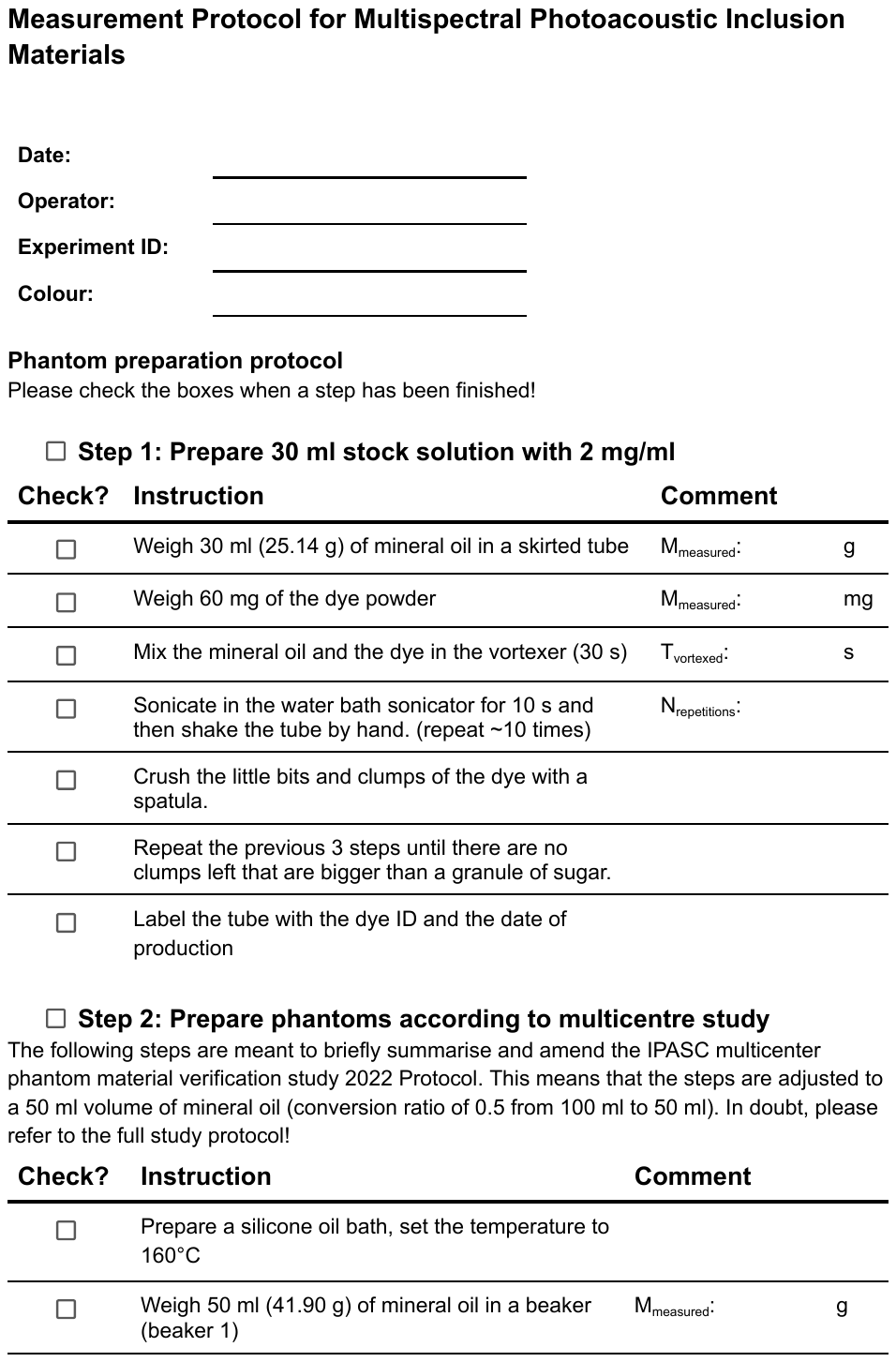}

\includegraphics[width=0.93\textwidth,page=2]{supp_figures/Baseline_BXX_Multispectral_Phantoms.pdf}

\section{Supplementary Figures}

\begin{figure}[H]
    \centering
    \includegraphics[width=\textwidth]{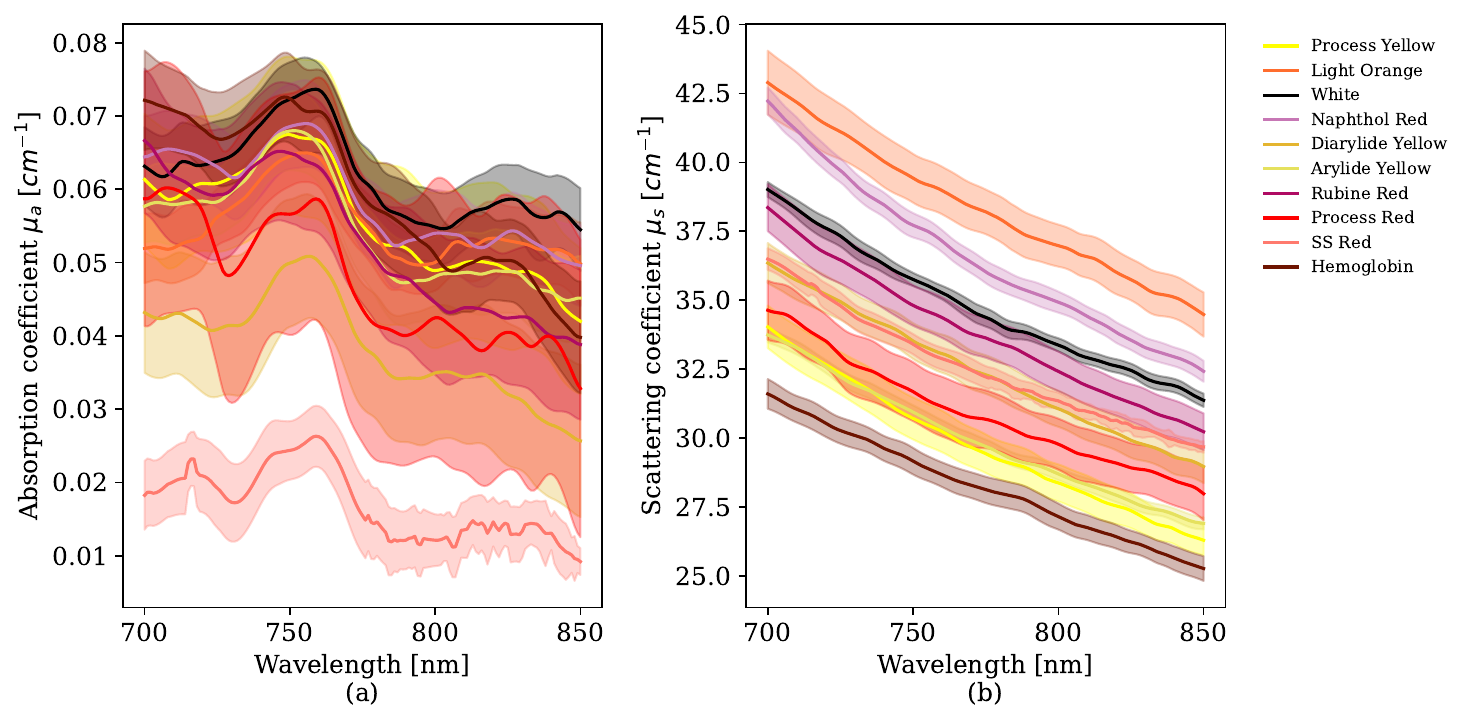}
    \caption{\textbf{Optical properties of investigated dyes with low absorption.} In total, 26 dyes were examined for their optical absorption (a) and scattering (b) in a copolymer-in-oil solution using a double-integrating sphere setup. 10 dyes were identified with relatively low absorption coefficient: Process Yellow, Light Orange, White, Naphthol Red, Diarylide Yellow, Rubine Red, Process Red, SS Red, Hemoglobin. For manufacturers and product names, refer to Table S1. Bands around the measured spectra indicate the standard deviation across the 12 measurement points for each optical sample slab.
    }
\end{figure}

\begin{figure}[H]
    \centering
    \includegraphics[width=\textwidth]{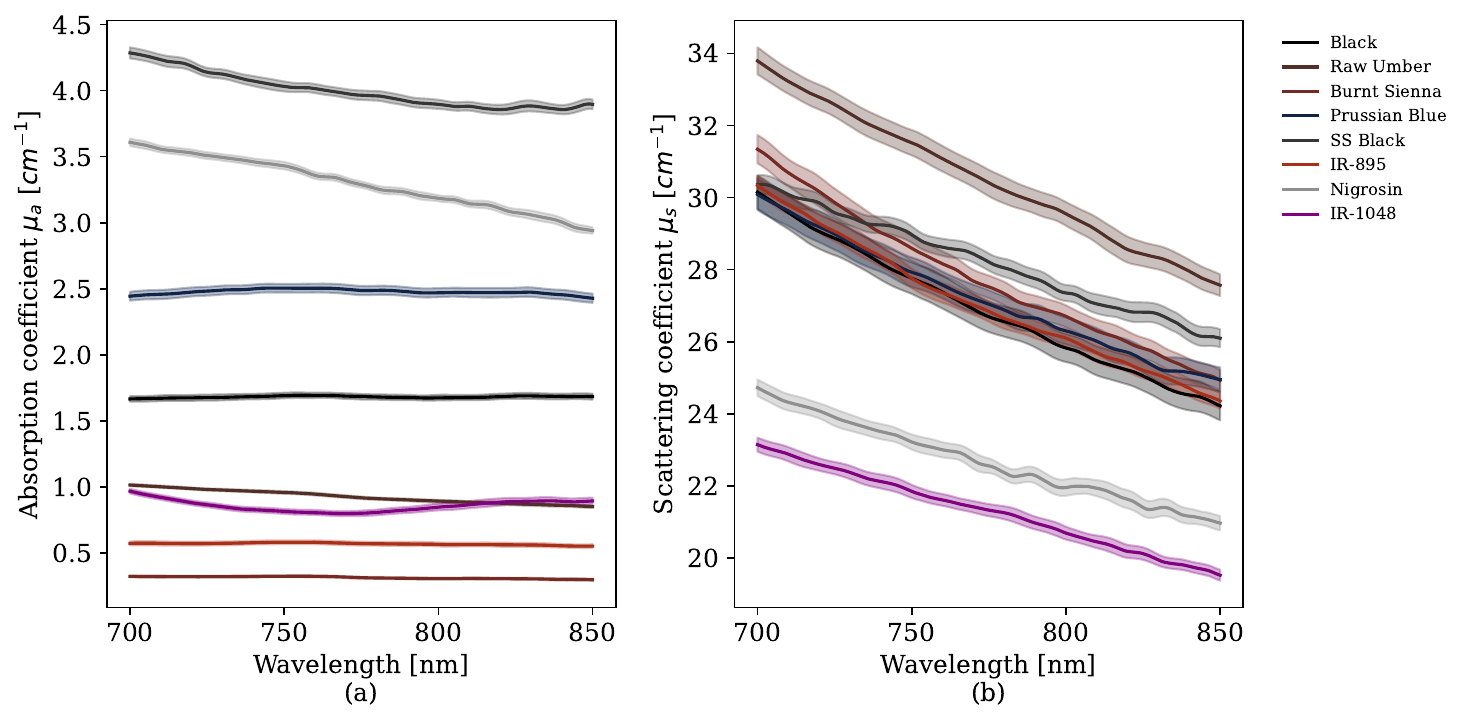}
    \caption{\textbf{Optical properties of investigated dyes with high absorption.} In total, 26 dyes were examined for their optical absorption (a) and scattering (b) in a copolymer-in-oil solution using a double-integrating sphere setup. 8 dyes were identified with relatively high absorption coefficient: Black, Raw Umber, Burnt Sienna, Prussian Blue, SS Black, IR-895, Nigrosin, IR-1048. For manufacturers and product names, refer to Table S1. Bands around the measured spectra indicate the standard deviation across the 12 measurement points for each optical sample slab.
    }
\end{figure}

\begin{figure}[H]
    \centering
    \includegraphics[width=\textwidth]{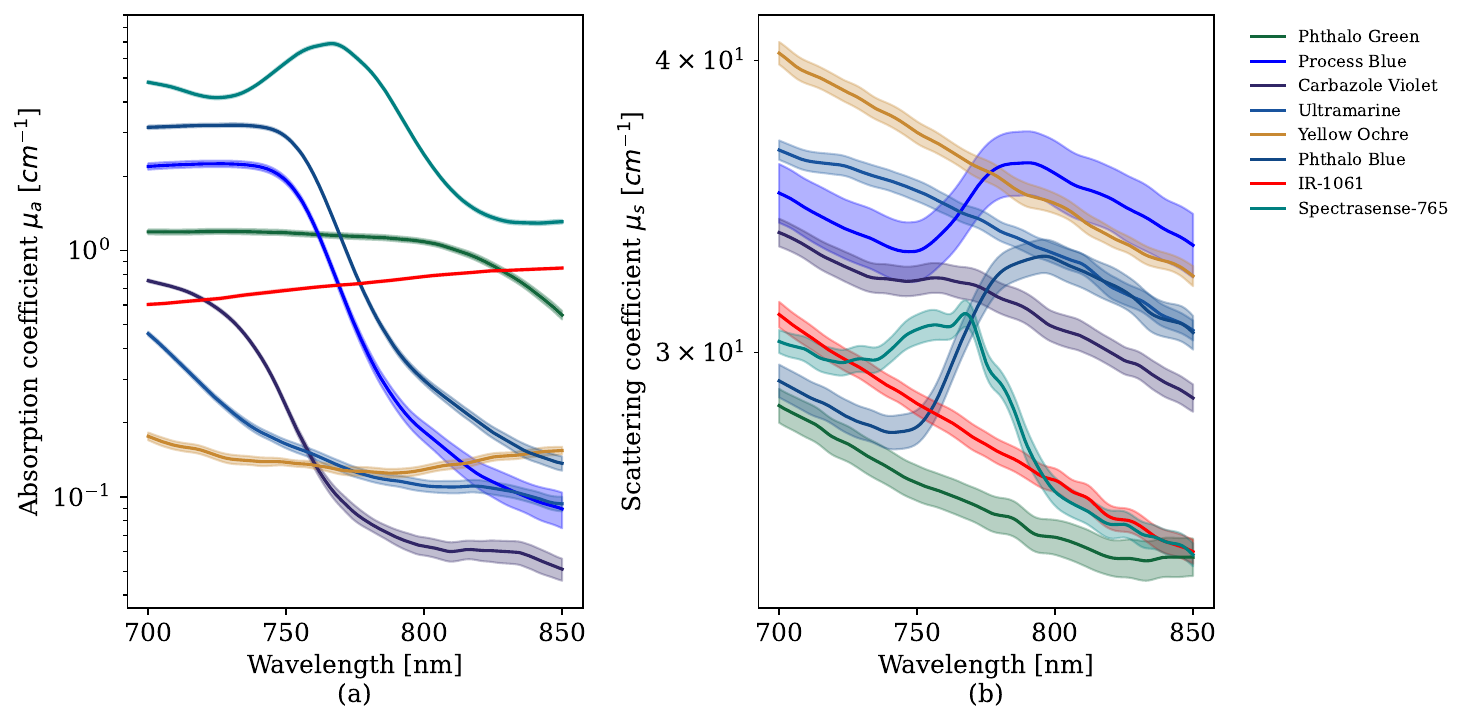}
    \caption{\textbf{Optical properties of investigated dyes with high range.} In total, 26 dyes were examined for their optical absorption (a) and scattering (b) in a copolymer-in-oil solution using a double-integrating sphere setup. 8 dyes were identified with a relatively high range of absorption coefficient: Phthalo Green, Process Blue, Carbazole Violet, Ultramarine, Yellow Ochre, Phthalo Blue, IR-1061, Spectrasense-765. For manufacturers and product names, refer to Table S1. Bands around the measured spectra indicate the standard deviation across the 12 measurement points for each optical sample slab.
    }
\end{figure}

\begin{figure}[H]
    \centering
    \includegraphics[width=\textwidth]{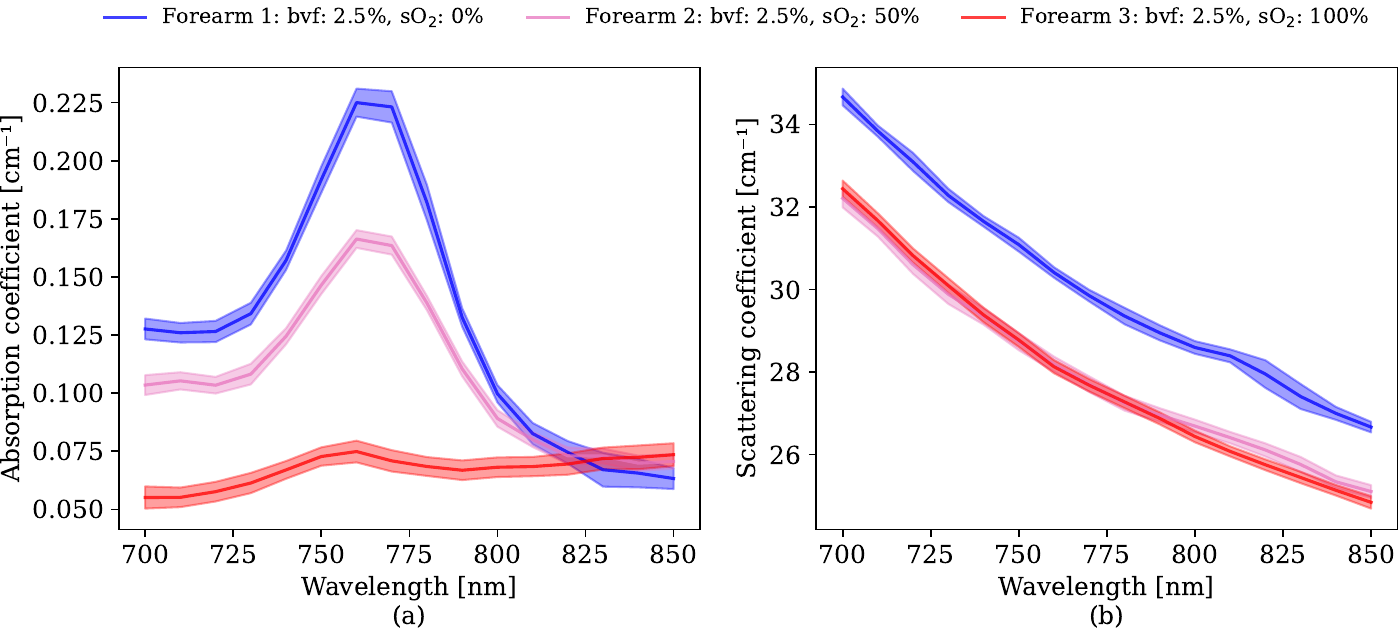}
    \caption{\textbf{Optical absorption (a) and scattering (a) coefficients for forearms 1-3.} For scattering anisotropy (g), a value of g=0.7 was assumed.
    }
\end{figure}

\begin{figure}[H]
    \centering
    \includegraphics[width=\textwidth]{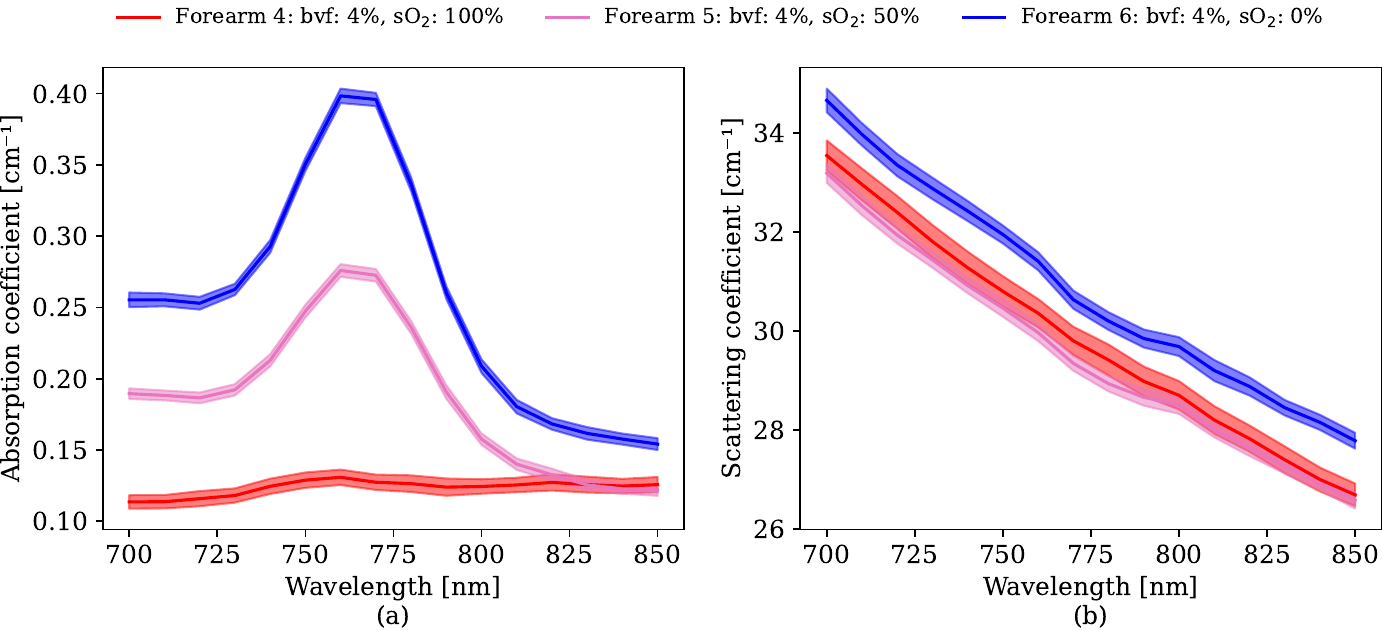}
    \caption{\textbf{Optical absorption (a) and scattering (a) coefficients for forearms 4-6.} For scattering anisotropy (g), a value of g=0.7 was assumed.
    }
\end{figure}

\begin{figure}[H]
    \centering
    \includegraphics[width=\textwidth]{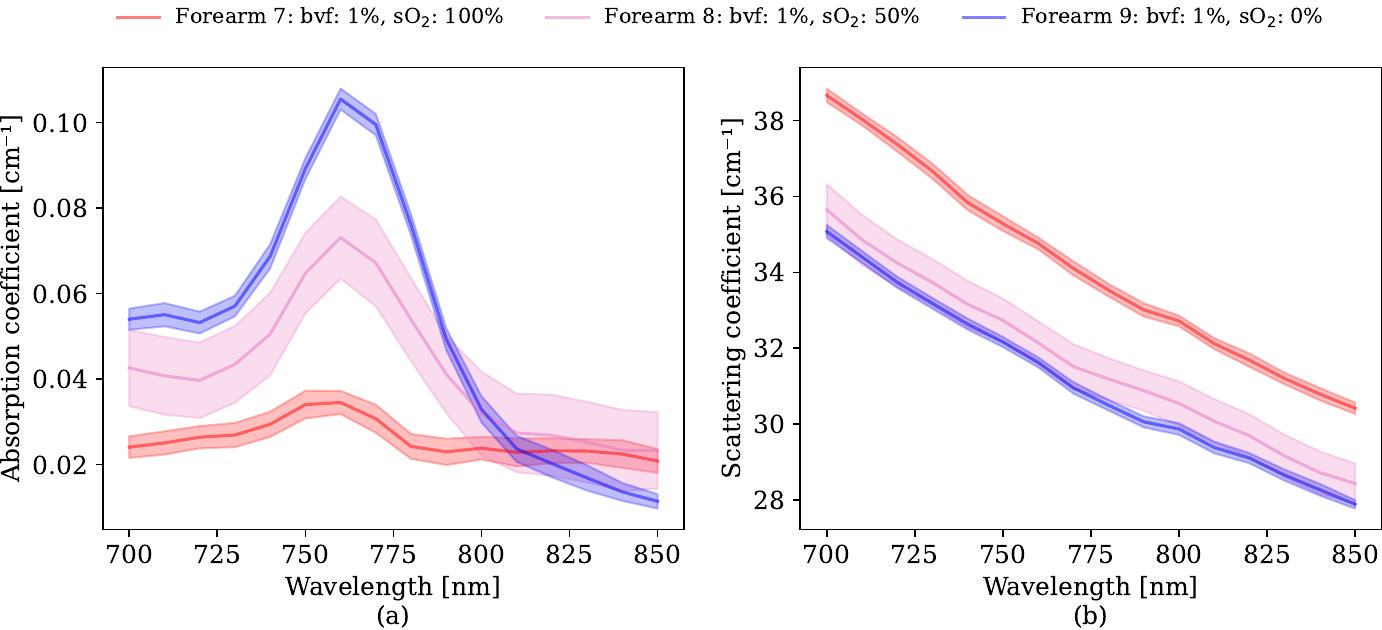}
    \caption{\textbf{Optical absorption (a) and scattering (a) coefficients for forearms 7-9.} For scattering anisotropy (g), a value of g=0.7 was assumed.
    }
\end{figure}

\begin{figure}[H]
    \centering
    \includegraphics[width=\textwidth]{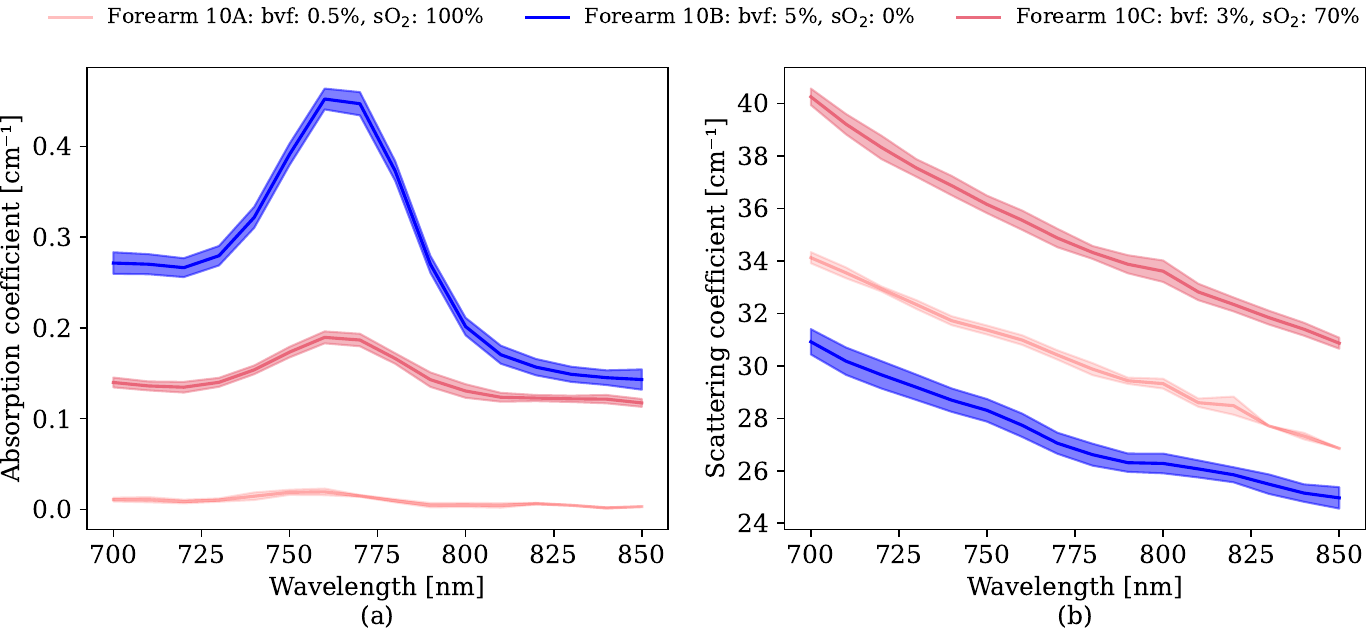}
    \caption{\textbf{Optical absorption (a) and scattering (b) coefficients for all three background compositions of forearm 10 (10A, 10B, and 10C).} For scattering anisotropy (g), a value of g=0.7 was assumed.
    }
\end{figure}

\begin{figure}[H]
    \centering
    \includegraphics[width=\textwidth]{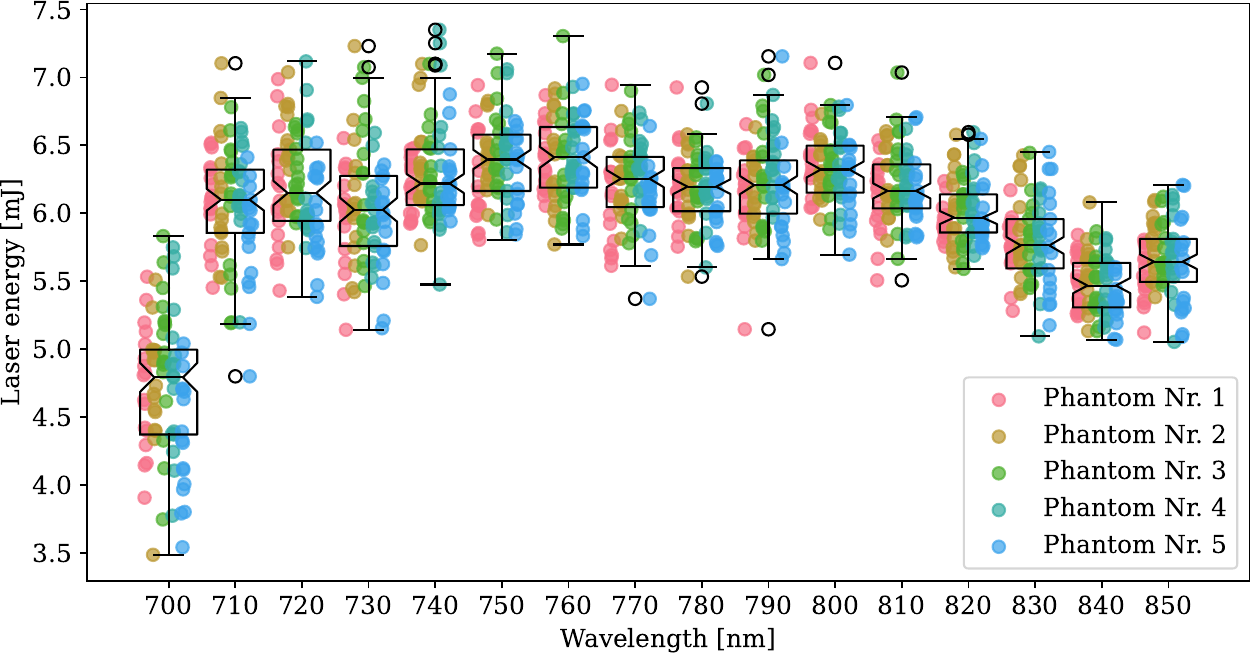}
    \caption{\textbf{Measured laser energies in mJ for each of the example phantoms' 27 photoacoustic images.}
    }
\end{figure}

\begin{figure}[!ht]
    \centering
    \includegraphics[width=0.6\textwidth]{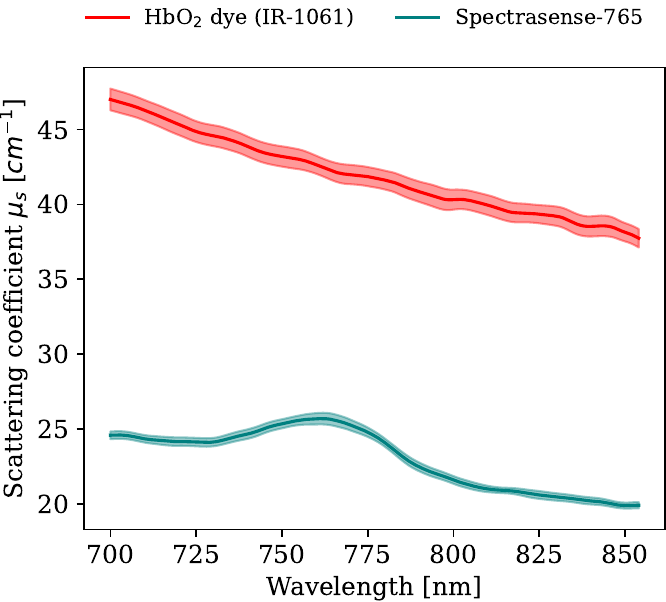}
    \caption{\textbf{Scattering spectra of IR-1061 and Spectrasense-765.} Solid lines indicate the measured scattering $\mu_s$ spectra of the proxy dyes. Bands around the measured spectra indicate the standard deviation across the 12 measurement points for each optical sample slab.
    }
\end{figure}

\begin{figure}[H]
    \centering
    \includegraphics[width=\textwidth]{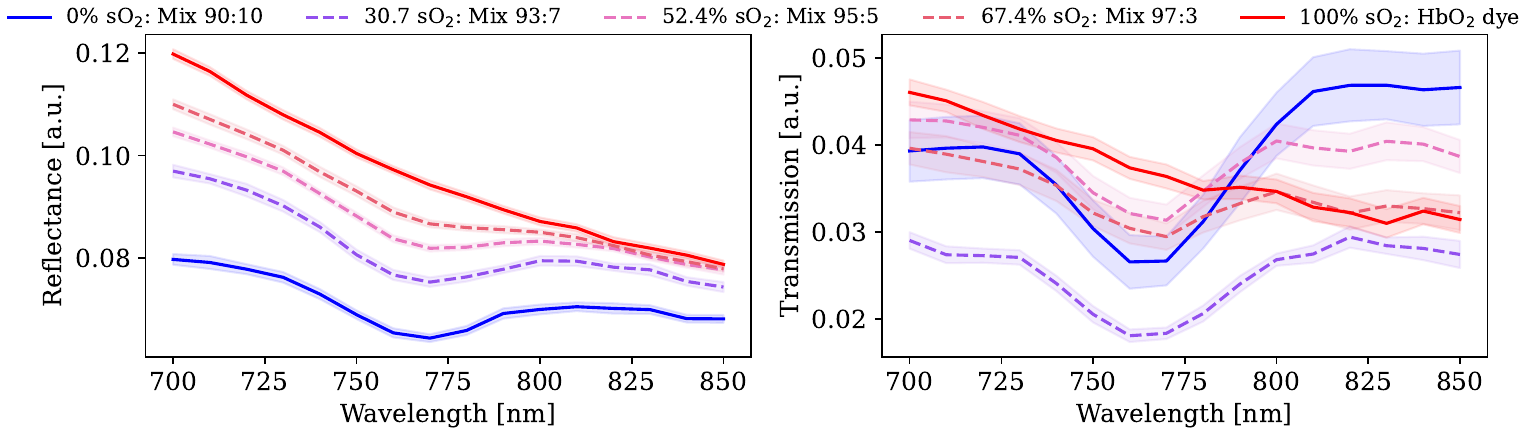}
    \caption{\textbf{Raw reflectance and transmission spectra from the DIS system for the five oxygen saturation levels were used for forearm phantom fabrication.} Based on IR-1061 and Spectrasense-765, five oxygen saturation levels (in \%) were derived with the respective mixture ratios of 100:0, 97:3, 95:5, 93:7, 90:10. (a) and (b) represent the double integrating sphere measurements of reflectance and transmission, respectively. Solid lines are the spectra that are used as endmembers (0\% oxygen saturation and 100\% oxygen saturation) for linear spectral unmixing (LSU). Dashed lines represent the intermediate levels and the corresponding percentages in the legend are the LSU results when using the solid lines as endmembers. Bands around the spectra indicate the standard deviation across the 12 measurement points for each optical sample slab.
    }
\end{figure}

\begin{figure}[H]
    \centering
	\begin{subfigure}[t]{0.49\textwidth}
		\centering
		\includegraphics[width=1\textwidth]{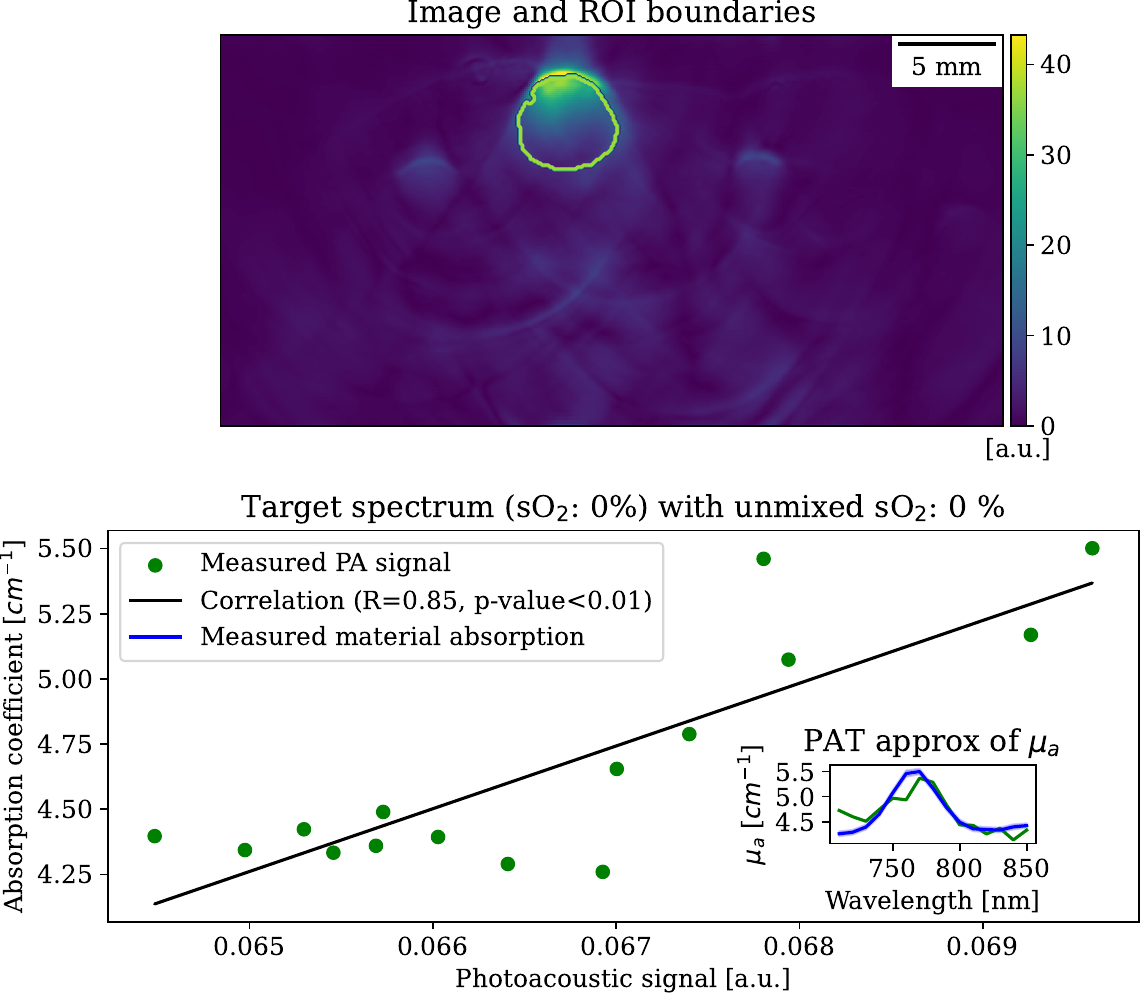}
		\caption{}	
	\end{subfigure}
	\begin{subfigure}[t]{0.49\textwidth}
		\centering
		\includegraphics[width=1\textwidth]{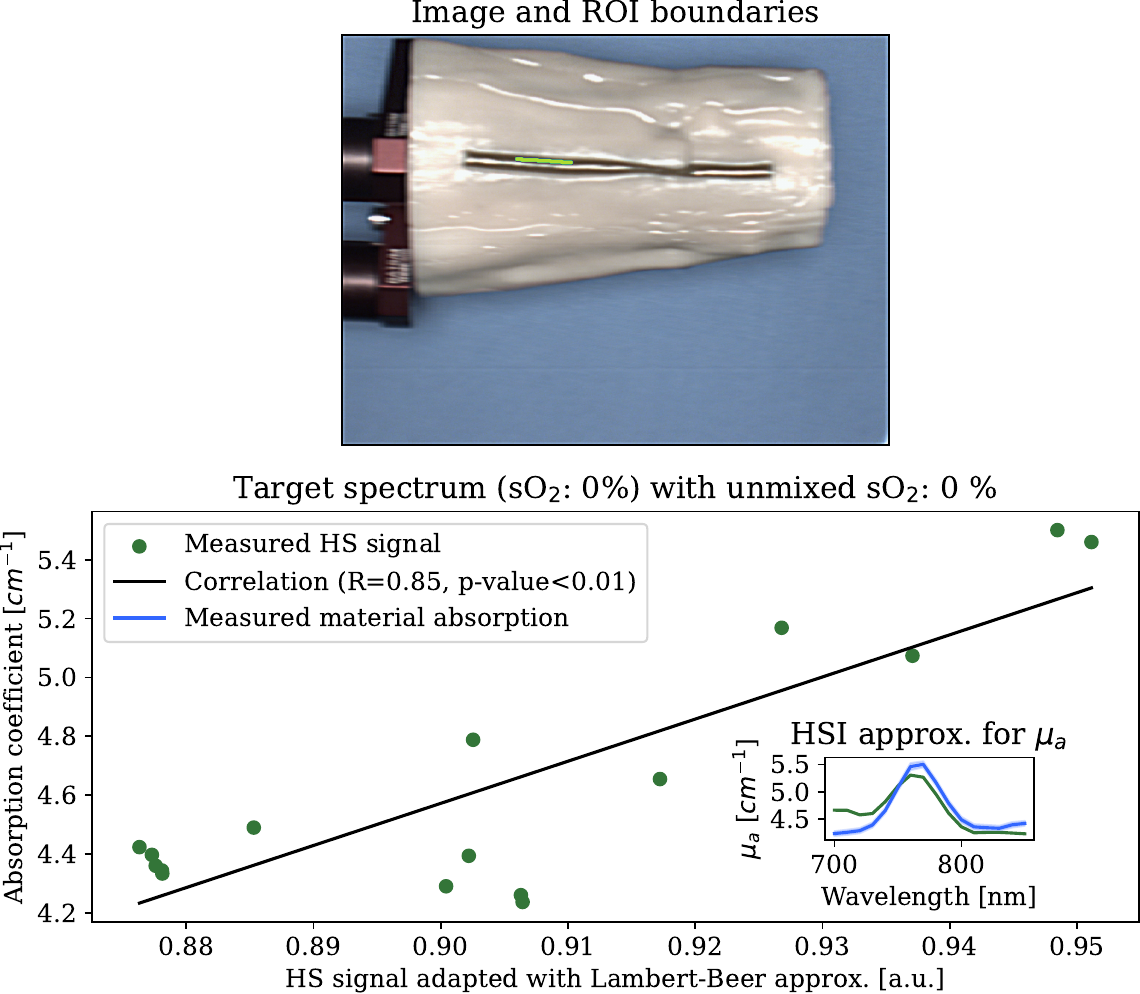}
		\caption{}
	\end{subfigure}
    \caption{
    \textbf{Examples of signal correlation of the 0\% oxygen saturation superficial vessel for both (a) photoacoustic (PA) tomography (PAT) and (b) hyperspectral (HS) imaging (HSI).} Top pictures show (a) PA image with region of interest (ROI) boundary (green outline) and (b) RGB- (red blue green) rendered HS image with its ROI boundary. Lower pictures show the correlation of measured spectra (green dots as explained in sec. 3 of the main paper) with the measured absorption coefficients. Black solid lines represent the resulting linear regression function with corresponding R values. The inset plots show the measured absorptions (blue) and estimated absorptions (green, using the correlation function) as qualitative confirmation.
    }
\end{figure}

\begin{figure}[H]
    \centering
	\begin{subfigure}[t]{0.49\textwidth}
		\centering
		\includegraphics[width=1\textwidth]{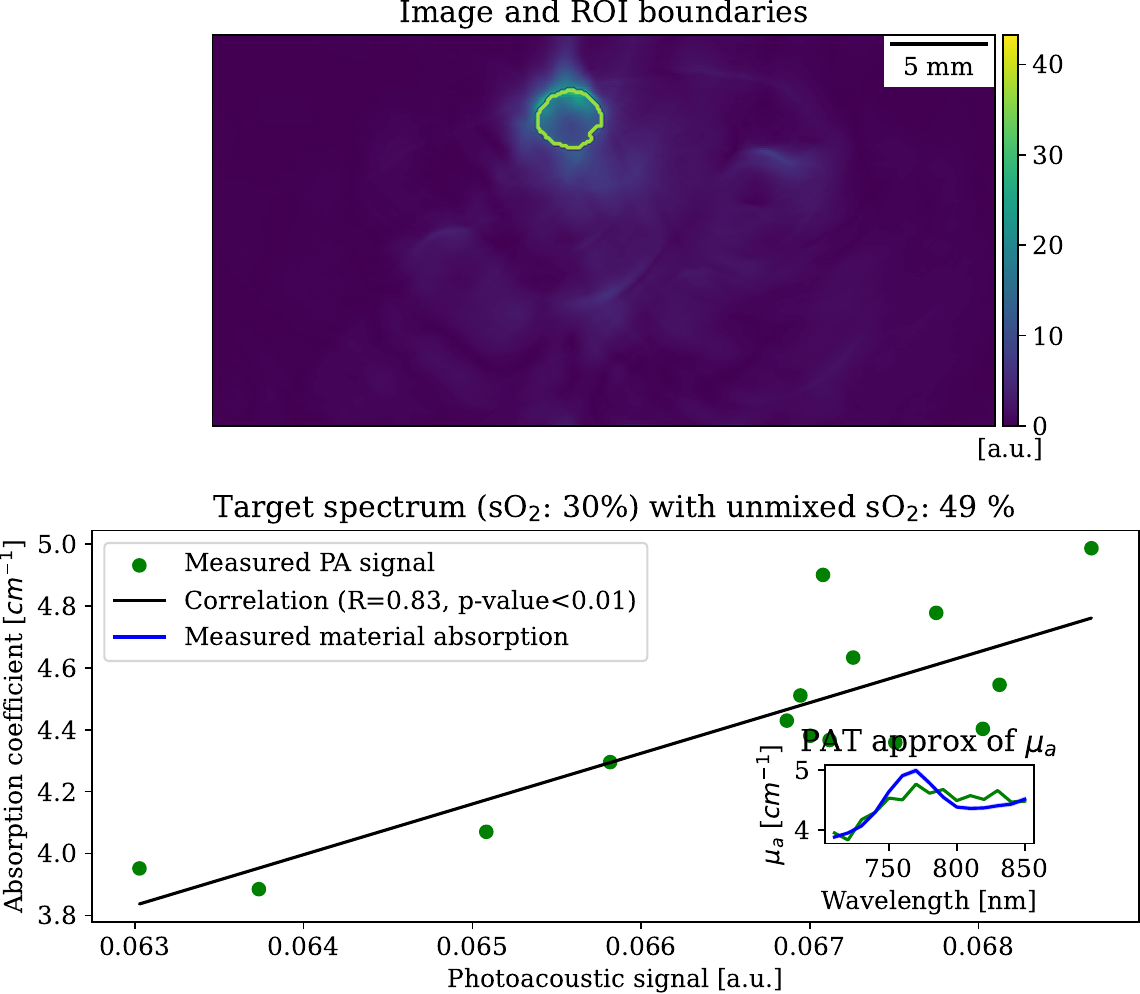}
		\caption{}		
	\end{subfigure}
	\begin{subfigure}[t]{0.49\textwidth}
		\centering
		\includegraphics[width=1\textwidth]{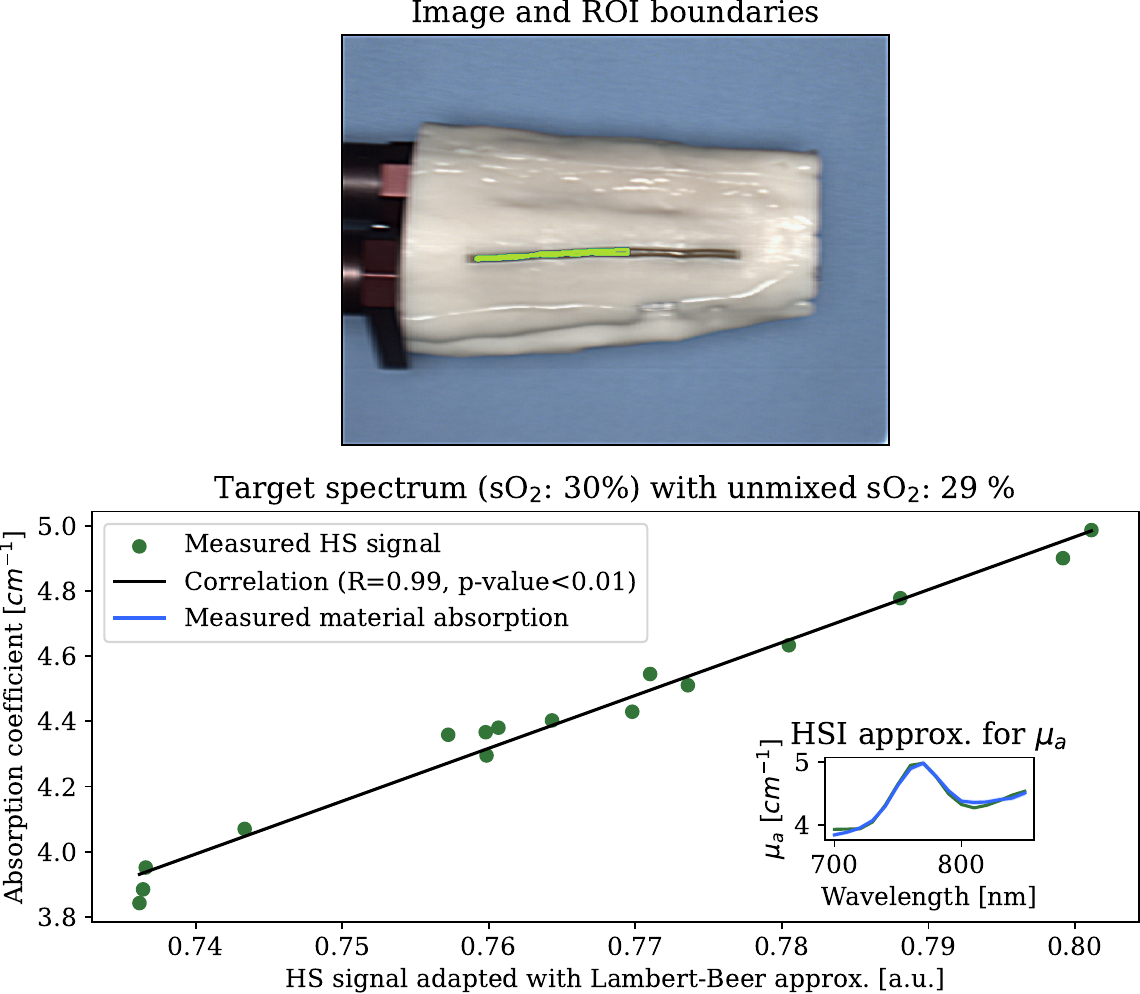}
		\caption{}
	\end{subfigure}
    \caption{
    \textbf{Examples of signal correlation of the 30\% oxygen saturation superficial vessel for both (a) photoacoustic (PA) tomography (PAT) and (b) hyperspectral (HS) imaging (HSI).} Top pictures show (a) PA image with region of interest (ROI) boundary (green outline) and (b) RGB- (red blue green) rendered HS image with its ROI boundary. Lower pictures show the correlation of measured spectra (green dots as explained in sec. 3 of the main paper) with the measured absorption coefficients. Black solid lines represent the resulting linear regression function with corresponding R values. The inset plots show the measured absorptions (blue) and estimated absorptions (green, using the correlation function) as qualitative confirmation.
    }
\end{figure}

\begin{figure}[H]
    \centering
	\begin{subfigure}[t]{0.49\textwidth}
		\centering
		\includegraphics[width=1\textwidth]{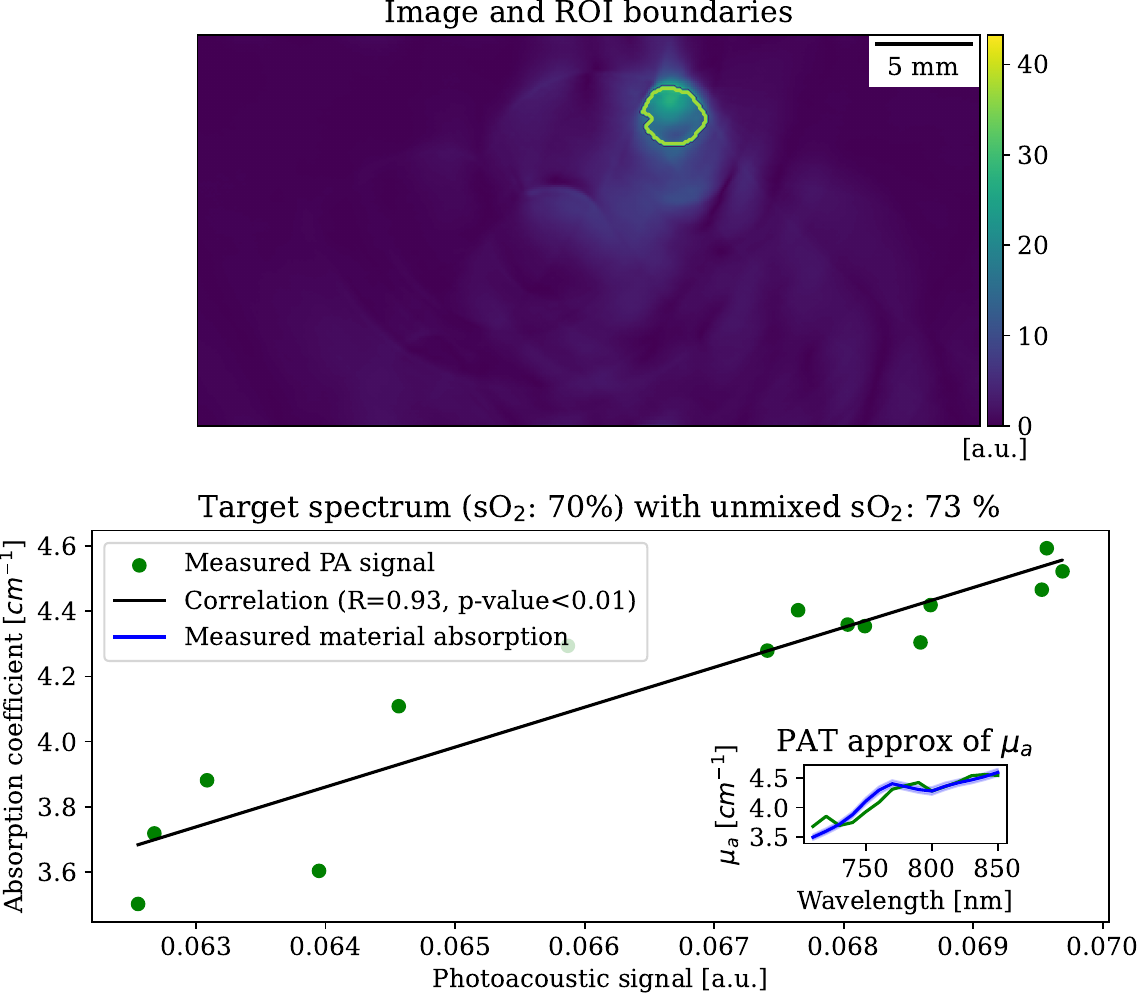}
		\caption{}		
	\end{subfigure}
	\begin{subfigure}[t]{0.49\textwidth}
		\centering
		\includegraphics[width=1\textwidth]{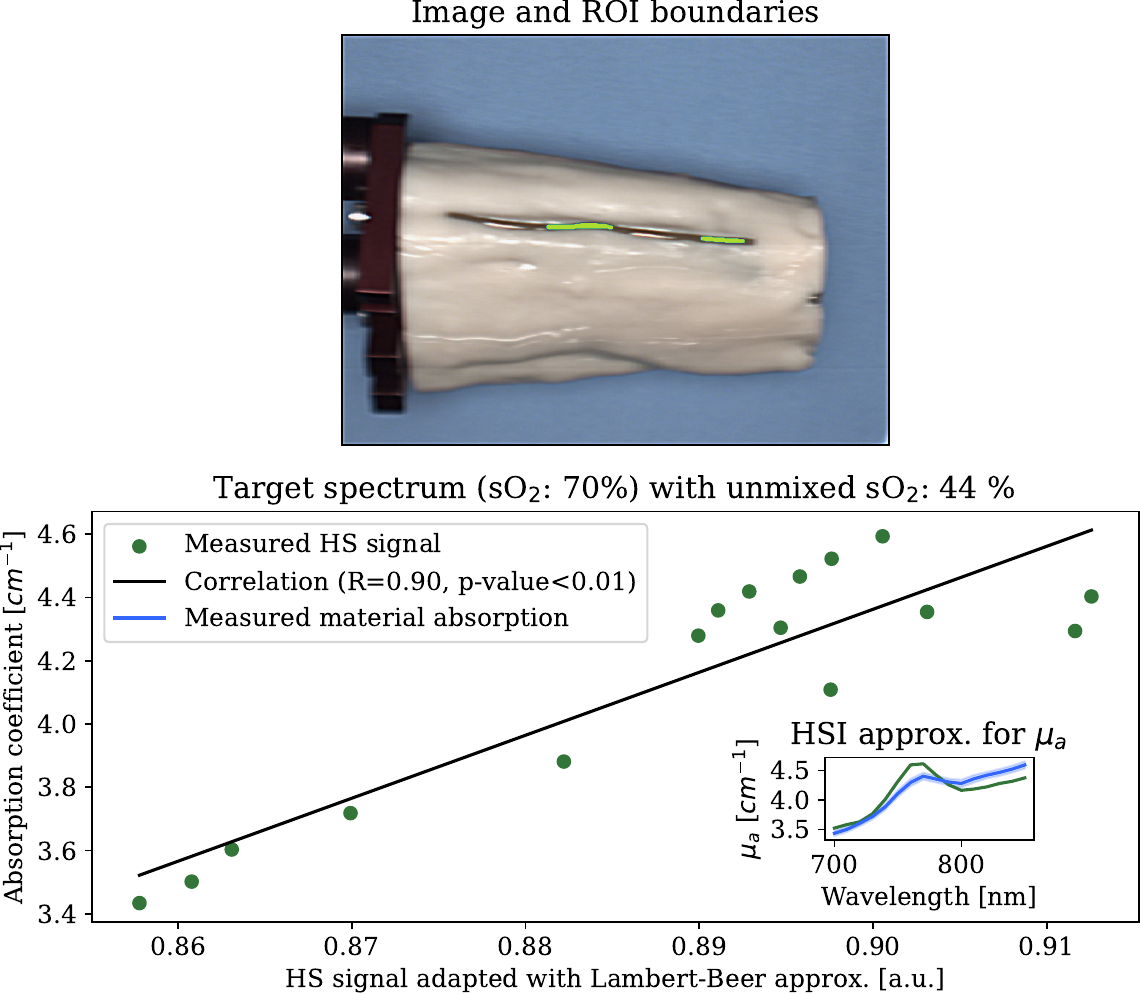}
		\caption{}
	\end{subfigure}
    \caption{
    \textbf{Examples of signal correlation of the 70\% oxygen saturation superficial vessel for both (a) photoacoustic (PA) tomography (PAT) and (b) hyperspectral (HS) imaging (HSI).} Top pictures show (a) PA image with region of interest (ROI) boundary (green outline) and (b) RGB- (red blue green) rendered HS image with its ROI boundary. Lower pictures show the correlation of measured spectra (green dots as explained in sec. 3 of the main paper) with the measured absorption coefficients. Black solid lines represent the resulting linear regression function with corresponding R values. The inset plots show the measured absorptions (blue) and estimated absorptions (green, using the correlation function) as qualitative confirmation.
    }
\end{figure}

\begin{figure}[H]
    \centering
	\begin{subfigure}[t]{0.49\textwidth}
		\centering
		\includegraphics[width=1\textwidth]{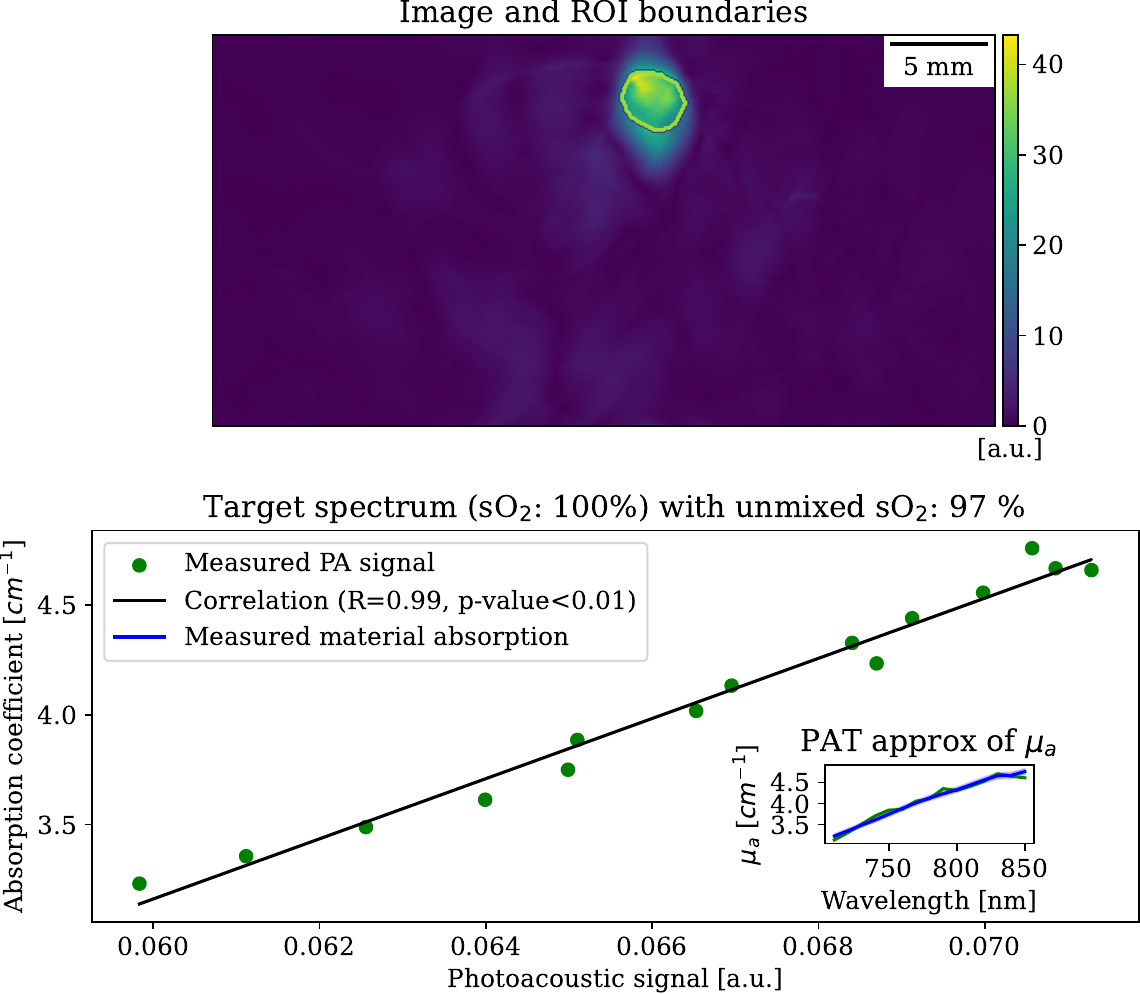}
		\caption{}	
	\end{subfigure}
	\begin{subfigure}[t]{0.49\textwidth}
		\centering
		\includegraphics[width=1\textwidth]{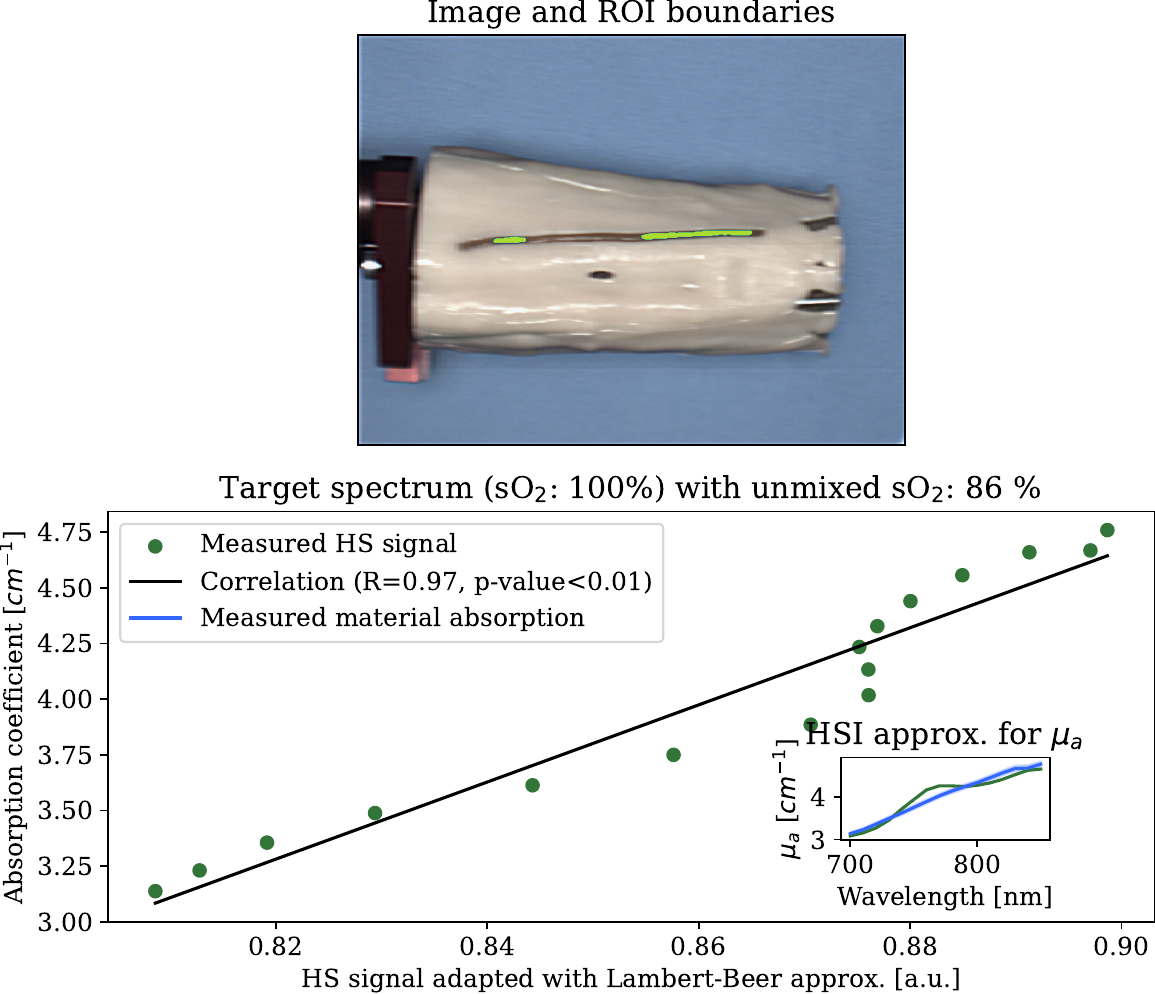}
		\caption{}
	\end{subfigure}
    \caption{
    \textbf{Examples of signal correlation of the 100\% oxygen saturation superficial vessel for both (a) photoacoustic (PA) tomography (PAT) and (b) hyperspectral (HS) imaging (HSI).} Top pictures show (a) PA image with region of interest (ROI) boundary (green outline) and (b) RGB- (red blue green) rendered HS image with its ROI boundary. Lower pictures show the correlation of measured spectra (green dots as explained in sec. 3 of the main paper) with the measured absorption coefficients. Black solid lines represent the resulting linear regression function with corresponding R values. The inset plots show the measured absorptions (blue) and estimated absorptions (green, using the correlation function) as qualitative confirmation.
    }
\end{figure}

\section{Phantom quality assurance}
\subsection{Air bubble investigation within phantoms}

Our forearm phantoms are considerably larger than many conventional tissue phantoms, making it difficult to achieve uniform fabrication without entrapping air. Indeed, in our experiments, we frequently observed air bubbles within the phantoms. To assess how these inclusions might affect the photoacoustic signal, we performed a small simulation study. First, we segmented example forearms (including visible air bubbles) from ultrasound images. We then simulated photoacoustic images for two conditions, one with and one without the air bubbles, and qualitatively compared the resulting vessel spectra.

\begin{figure}[H]
    \centering
    \includegraphics[width=0.7\textwidth]{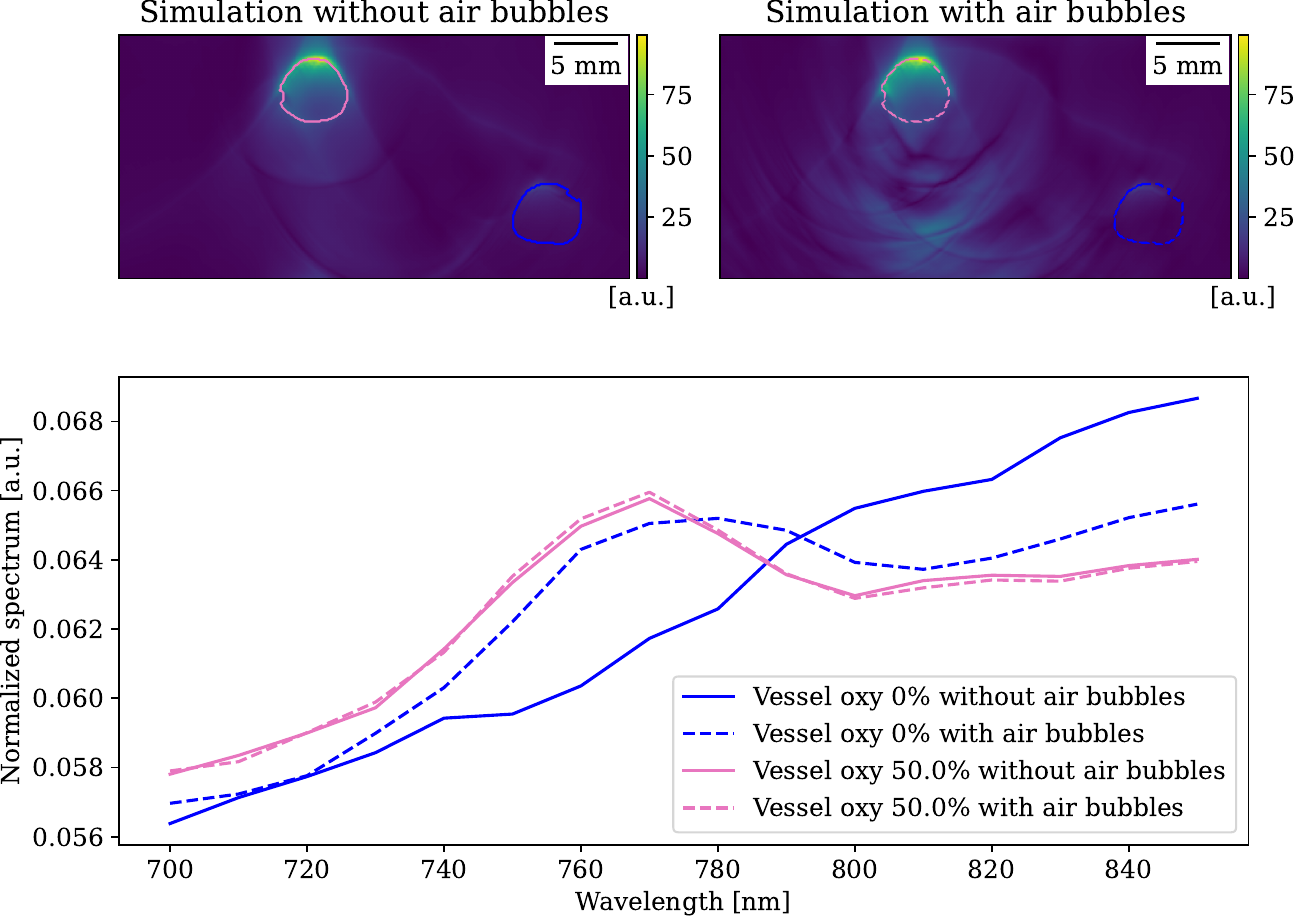}
    \caption{\textbf{Image and spectrum comparison of simulations with and without air bubbles.} Photoacoustic images of the forearm phantom 1 were simulated with (top right) and without (top left) air bubbles. Images are shown at 700\,nm. Two vessels with different oxygen saturations are marked in the images and their mean spectra are plotted below. The solid lines represent the spectra without air bubbles and the dashed lines represent the spectra with air bubbles.
    }
    \label{fig:supp:air1}
\end{figure}

\begin{figure}[H]
    \centering
    \includegraphics[width=0.7\textwidth]{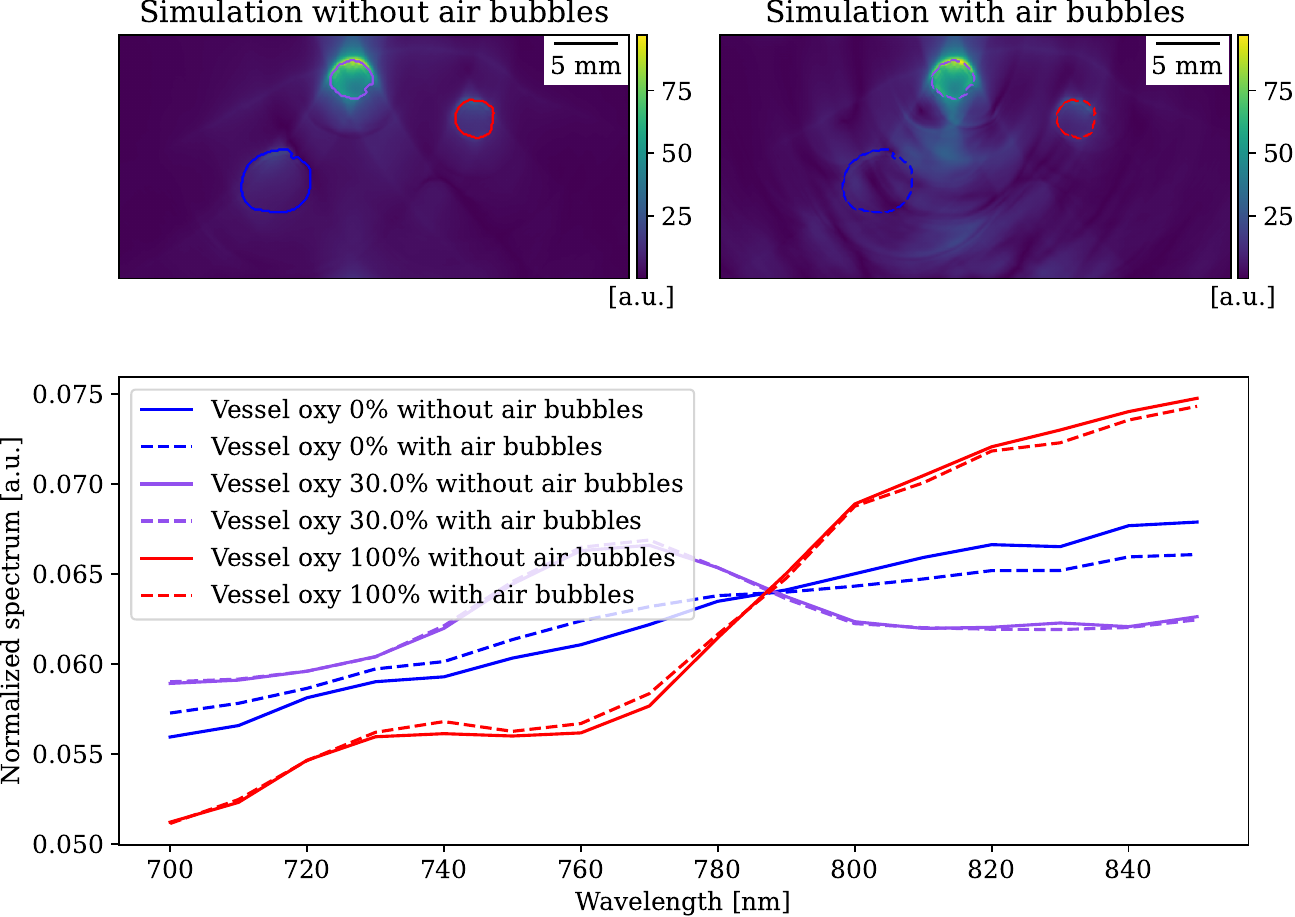}
    \caption{\textbf{Image and spectrum comparison of simulations with and without air bubbles.} Photoacoustic images of the forearm phantom 2 were simulated with (top right) and without (top left) air bubbles. Images are shown at 700\,nm. Three vessels with different oxygen saturations are marked in the images and their mean spectra are plotted below. The solid lines represent the spectra without air bubbles and the dashed lines represent the spectra with air bubbles.
    }
    \label{fig:supp:air2}
\end{figure}

\begin{figure}[H]
    \centering
    \includegraphics[width=0.7\textwidth]{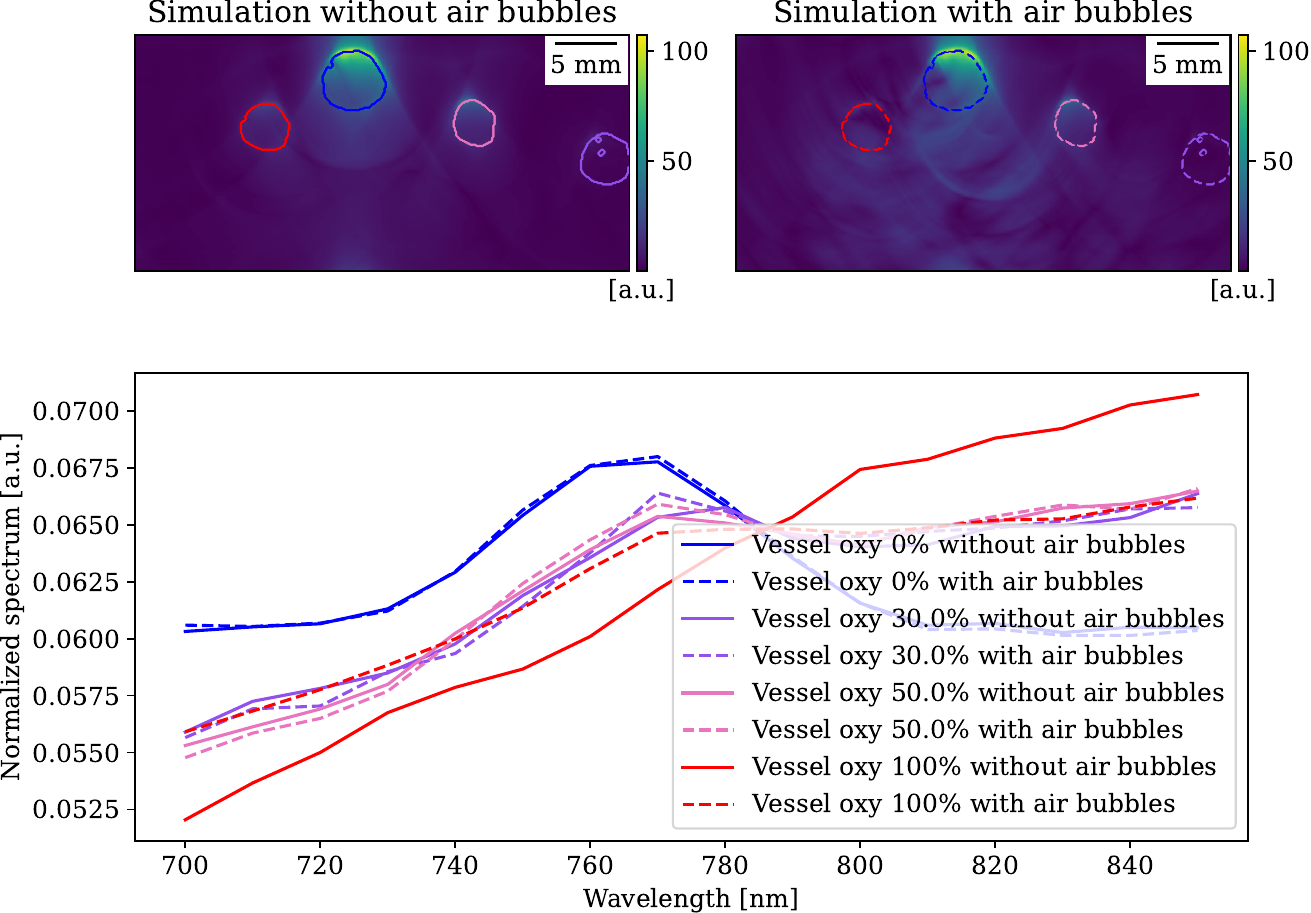}
    \caption{\textbf{Image and spectrum comparison of simulations with and without air bubbles.} Photoacoustic images of the forearm phantom 3 were simulated with (top right) and without (top left) air bubbles. Images are shown at 700\,nm. Four vessels with different oxygen saturations are marked in the images and their mean spectra are plotted below. The solid lines represent the spectra without air bubbles and the dashed lines represent the spectra with air bubbles. Small purple circles indicate air bubbles.
    }
    \label{fig:supp:air3}
\end{figure}

\begin{figure}[H]
    \centering
    \includegraphics[width=0.7\textwidth]{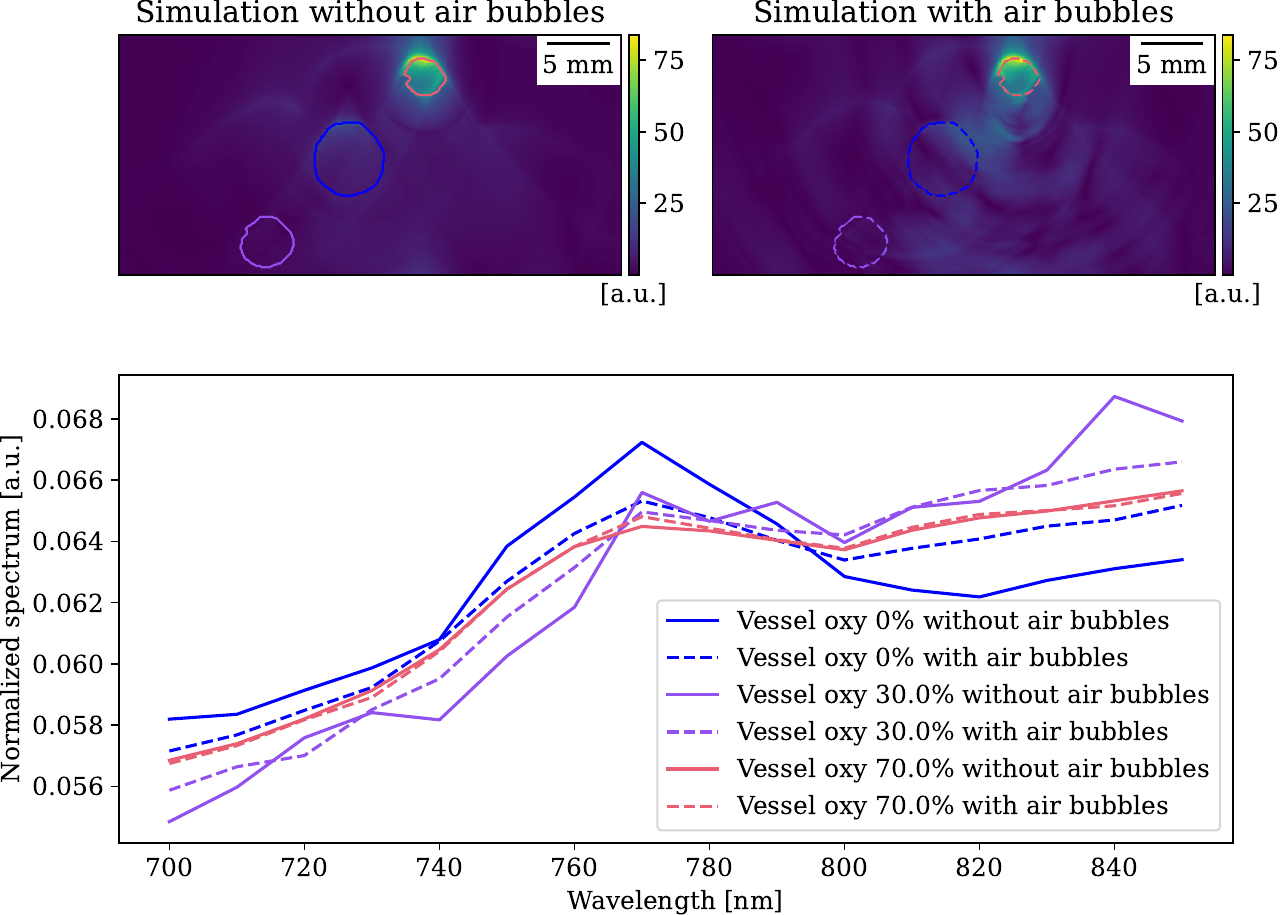}
    \caption{\textbf{Image and spectrum comparison of simulations with and without air bubbles.} Photoacoustic images of the forearm phantom 4 were simulated with (top right) and without (top left) air bubbles. Images are shown at 700\,nm. Three vessels with different oxygen saturations are marked in the images and their mean spectra are plotted below. The solid lines represent the spectra without air bubbles and the dashed lines represent the spectra with air bubbles.
    }
    \label{fig:supp:air4}
\end{figure}

\begin{figure}[H]
    \centering
    \includegraphics[width=0.7\textwidth]{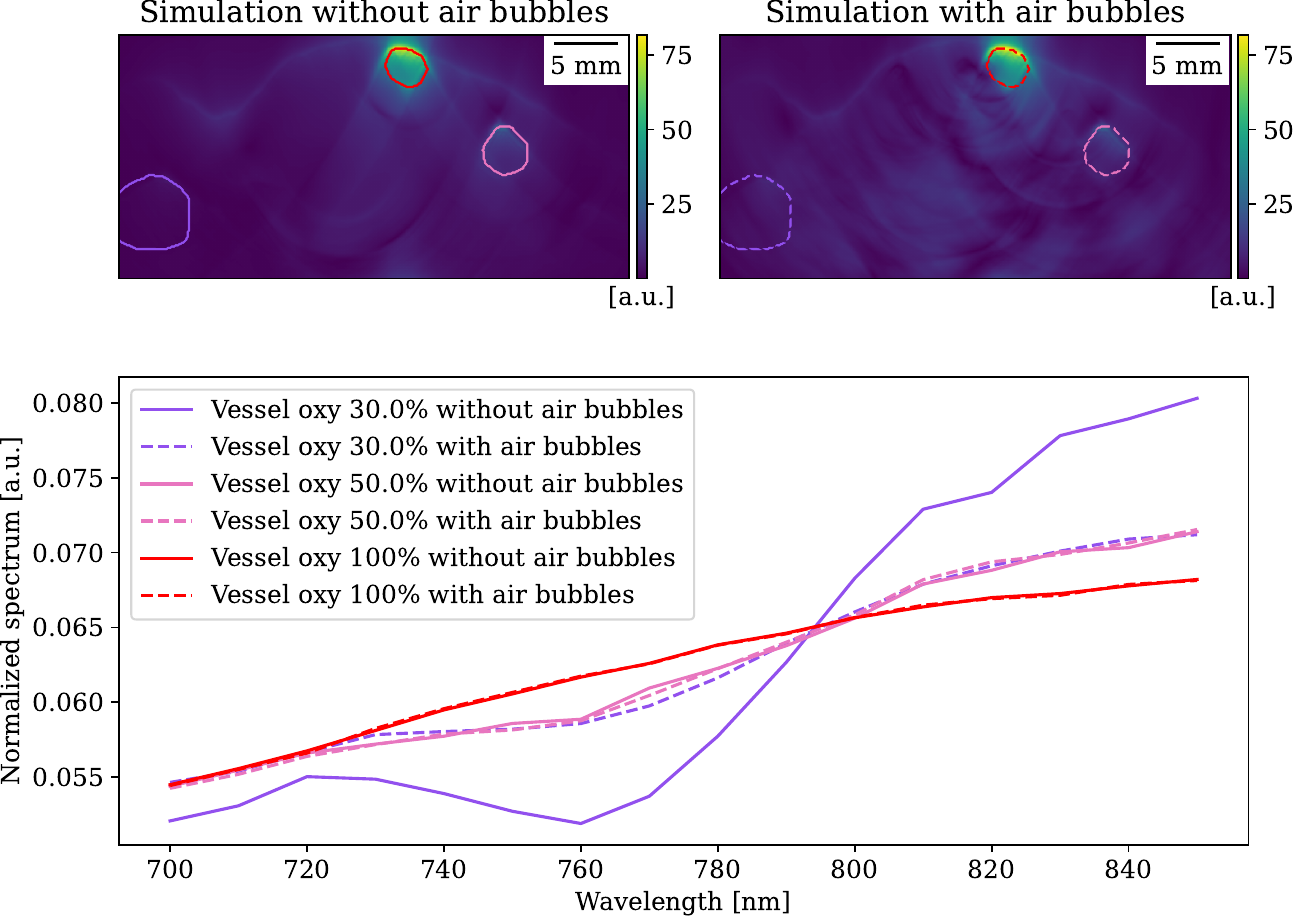}
    \caption{\textbf{Image and spectrum comparison of simulations with and without air bubbles.} Photoacoustic images of the forearm phantom 5 were simulated with (top right) and without (top left) air bubbles. Images are shown at 700\,nm. Three vessels with different oxygen saturations are marked in the images and their mean spectra are plotted below. The solid lines represent the spectra without air bubbles and the dashed lines represent the spectra with air bubbles.
    }
    \label{fig:supp:air5}
\end{figure}

Figure \ref{fig:supp:air1} illustrates two simulation outcomes for a digital twin of a forearm phantom, depicting scenarios with and without air bubbles. Two vessels, along with their respective oxygen saturation levels, are highlighted, and the mean spectra within each vessel are plotted for both conditions. In superficial vessels, the spectra agreed closely across both models, whereas deeper vessels showed noticeable distortion in the presence of air bubbles. Similar findings were observed in additional forearm phantoms (see Figures \ref{fig:supp:air2}, \ref{fig:supp:air3}, \ref{fig:supp:air4}, and \ref{fig:supp:air5}). Overall, vessels with centres located deeper than approximately 1.5 cm from the surface appear susceptible to reverberation artefacts caused by the embedded air bubbles.

\subsection{Speed of sound investigation}

We assumed a speed of sound of approximately 1470\,ms$^{-1}$ in our forearm phantoms. To evaluate whether this approximation is reasonable or significantly off, we conducted a small simulation study using segmentations from the previously described air-bubble investigation. First, we simulated photoacoustic images of the phantoms at\,1470\,ms$^{-1}$ and then repeated the process with the same segmentations at $\pm$50\,ms$^{-1}$ and $\pm$100\,ms$^{-1}$ from this baseline value.

Next, we manually segmented superficial vessels in each simulated image according to the previously described protocol, computed the segmented vessel area (A), and derived an approximate vessel diameter ($D$) from $D=2\sqrt{A/\pi}$. We compared these diameters to a reference diameter in the image $D_\text{im}$, calculated using the fabricated diameter ($D_\text{fab}$) and the ratio between the assumed phantom speed of sound ($sos_\text{ref}$) and the reconstruction speed of sound ($sos_\text{recon}$). Specifically, the reference diameter is given by $D_\text{im}=\frac{sos_\text{recon}}{sos_\text{ref}}D_\text{fab}$.

\begin{figure}[H]
    \centering
    \includegraphics[width=0.85\textwidth]{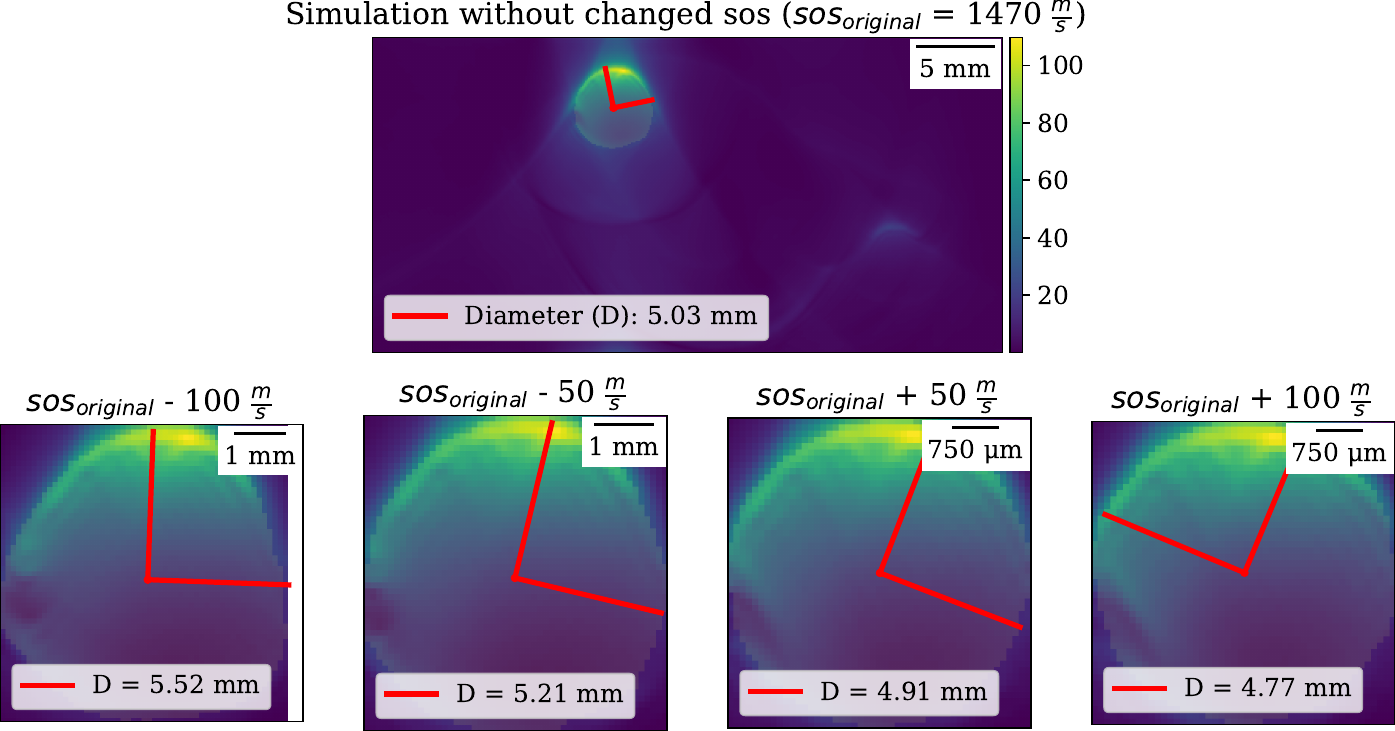}
    \caption{\textbf{Vessel diameter comparison for different speeds of sound.} Photoacoustic images of the forearm phantom 1 were simulated with a phantom material speed of sound of 1470\,ms$^{-1}$ (top) and then with speeds of sounds that differ from 1470\,ms$^{-1}$ by -100\,ms$^{-1}$, -50\,ms$^{-1}$, +50\,ms$^{-1}$, +100\,ms$^{-1}$ (from left to right). As the manual segmentations of the superficial vessel were not perfectly circular, the segmentation area A for all images was computed and their diameter $D$ was calculated via $D=2\sqrt{A/\pi}$. $D$ is indicated in all images by the major axes of the segmented areas.
    }
    \label{fig:supp:sos1}
\end{figure}

\begin{figure}[H]
    \centering
    \includegraphics[width=0.85\textwidth]{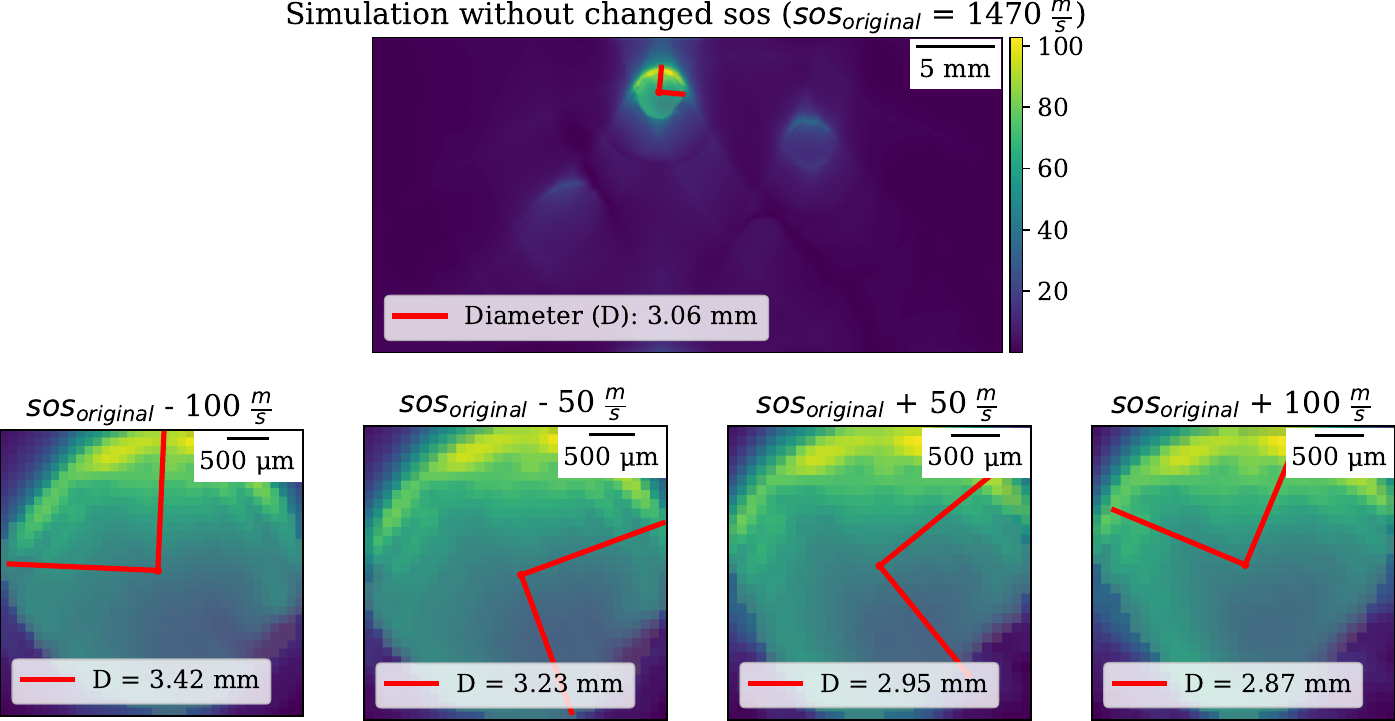}
    \caption{\textbf{Vessel diameter comparison for different speeds of sound.} Photoacoustic images of the forearm phantom 2 were simulated with a phantom material speed of sound of 1470\,ms$^{-1}$ (top) and then with speeds of sounds that differ from 1470\,ms$^{-1}$ by -100\,ms$^{-1}$, -50\,ms$^{-1}$, +50\,ms$^{-1}$, +100\,ms$^{-1}$ (from left to right). As the manual segmentations of the superficial vessel were not perfectly circular, the segmentation area A for all images was computed and their diameter $D$ was calculated via $D=2\sqrt{A/\pi}$. $D$ is indicated in all images by the major axes of the segmented areas.
    }
    \label{fig:supp:sos2}
\end{figure}

\begin{figure}[H]
    \centering
    \includegraphics[width=0.85\textwidth]{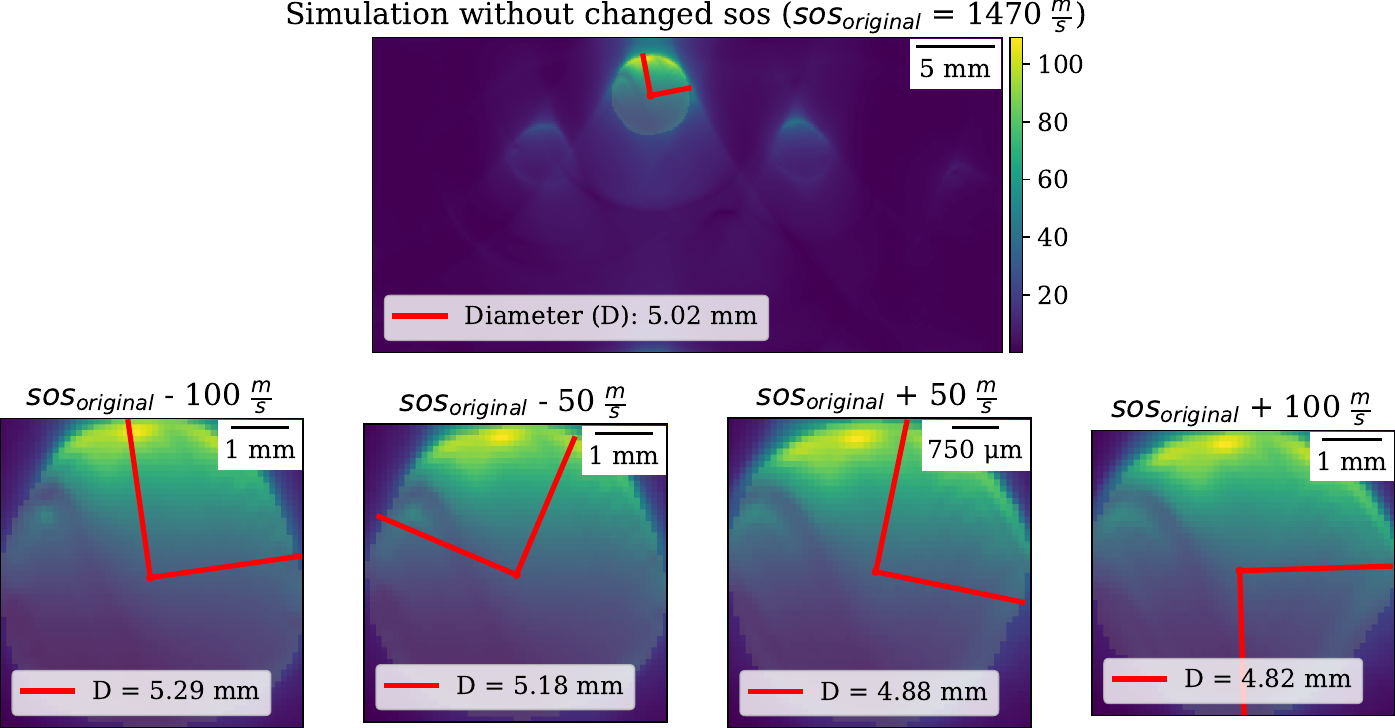}
    \caption{\textbf{Vessel diameter comparison for different speeds of sound.} Photoacoustic images of the forearm phantom 3 were simulated with a phantom material speed of sound of 1470\,ms$^{-1}$ (top) and then with speeds of sounds that differ from 1470\,ms$^{-1}$ by -100\,ms$^{-1}$, -50\,ms$^{-1}$, +50\,ms$^{-1}$, +100\,ms$^{-1}$ (from left to right). As the manual segmentations of the superficial vessel were not perfectly circular, the segmentation area A for all images was computed and their diameter $D$ was calculated via $D=2\sqrt{A/\pi}$. $D$ is indicated in all images by the major axes of the segmented areas.
    }
    \label{fig:supp:sos3}
\end{figure}

\begin{figure}[H]
    \centering
    \includegraphics[width=0.85\textwidth]{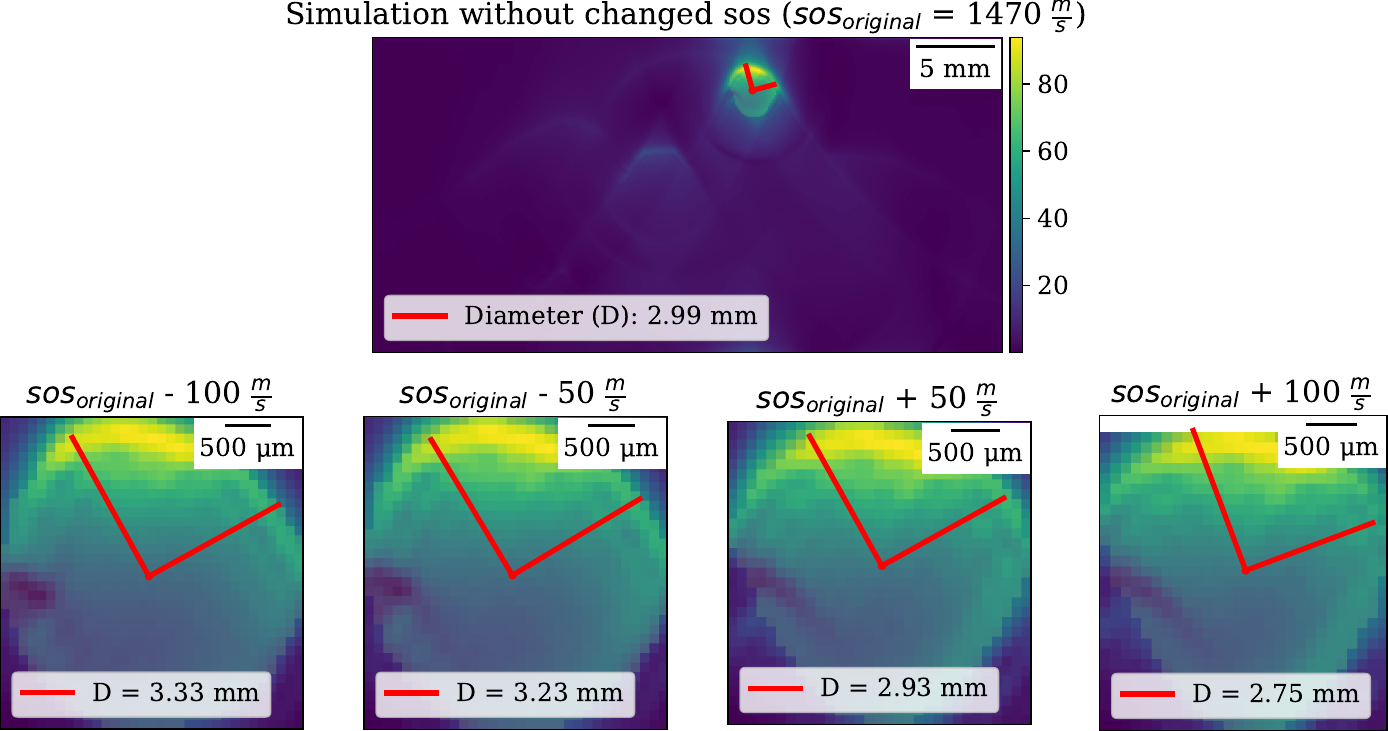}
    \caption{\textbf{Vessel diameter comparison for different speeds of sound.} Photoacoustic images of the forearm phantom 4 were simulated with a phantom material speed of sound of 1470\,ms$^{-1}$ (top) and then with speeds of sounds that differ from 1470\,ms$^{-1}$ by -100\,ms$^{-1}$, -50\,ms$^{-1}$, +50\,ms$^{-1}$, +100\,ms$^{-1}$ (from left to right). As the manual segmentations of the superficial vessel were not perfectly circular, the segmentation area A for all images was computed and their diameter $D$ was calculated via $D=2\sqrt{A/\pi}$. $D$ is indicated in all images by the major axes of the segmented areas.
    }
    \label{fig:supp:sos4}
\end{figure}

\begin{figure}[H]
    \centering
    \includegraphics[width=0.85\textwidth]{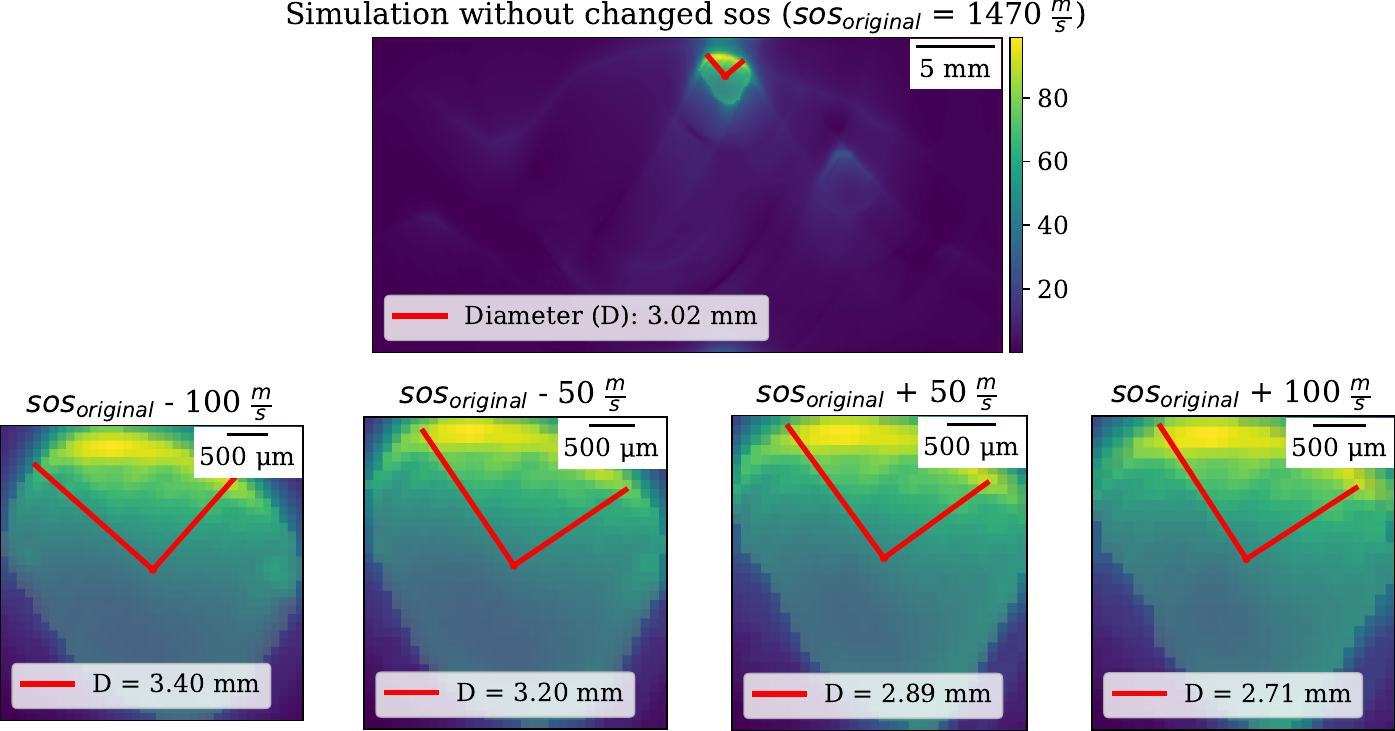}
    \caption{\textbf{Vessel diameter comparison for different speeds of sound.} Photoacoustic images of the forearm phantom 5 were simulated with a phantom material speed of sound of 1470\,ms$^{-1}$ (top) and then with speeds of sounds that differ from 1470\,ms$^{-1}$ by -100\,ms$^{-1}$, -50\,ms$^{-1}$, +50\,ms$^{-1}$, +100\,ms$^{-1}$ (from left to right). As the manual segmentations of the superficial vessel were not perfectly circular, the segmentation area A for all images was computed and their diameter $D$ was calculated via $D=2\sqrt{A/\pi}$. $D$ is indicated in all images by the major axes of the segmented areas.
    }
    \label{fig:supp:sos5}
\end{figure}

Figure \ref{fig:supp:sos1} shows the simulation results for a digital twin of a forearm phantom, assuming an sos of 1470\,ms$^{-1}$ and variations of $\pm$50 ms$^{-1}$ and $\pm$100 ms$^{-1}$ (for the other phantoms, cf. Figures \ref{fig:supp:sos2}, \ref{fig:supp:sos3}, \ref{fig:supp:sos4}, and \ref{fig:supp:sos5}. In Fig. \ref{fig:supp:sos1}, the fabricated vessel diameter was 5\,mm. Under the assumption of 1470\,ms$^{-1}$ with a reconstruction at 1497.4\,ms$^{-1}$, we would expect an imaged diameter of 5.09\,mm. For the other sos values of -100\,ms$^{-1}$, -50\,ms$^{-1}$, +50\,ms$^{-1}$, and +100\,ms$^{-1}$, the respective imaged diameters were 5.46\,mm, 5.27\,mm, 4.93\,mm, and 4.77\,mm. Given the simulated image resolution of 0.1\,mm, these findings align closely with theoretical expectations, suggesting that 1470\,ms$^{-1}$ is a sufficiently accurate estimate for our phantom material.

\section{Supplementary Tables}

\begin{table}[H]
    \centering
    \begin{tabular}{l l l}
		Dye name & Manufacturer/Vendor & Product name\\
        \hline
        Black & Cranfield (Jackson's art) & RCR2501860\\
        Process Blue & Cranfield (Jackson's art) & RCR25025291\\
        Process Red & Cranfield (Jackson's art) & RCR25063827\\
        Process Yellow & Cranfield (Jackson's art) & RCR25091884\\
        Arylide Yellow & Cranfield (Jackson's art) & RCR25091630\\
        Diarylide Yellow & Cranfield (Jackson's art) & RCR25091759\\
        Light Orange & Cranfield (Jackson's art) & RCR25091637\\
        Rubine Red & Cranfield (Jackson's art) & RCR25063254\\
        Naphthol Red & Cranfield (Jackson's art) & RCR50063266\\
        Carbazole Violet & Cranfield (Jackson's art) & RCR50071139\\
        Ultramarine & Cranfield (Jackson's art) & RCR50024283\\
        Prussian Blue & Cranfield (Jackson's art) & RCR25024309\\
        Phthalo Blue & Cranfield (Jackson's art) & RCR25024760\\
        Phthalo Green & Cranfield (Jackson's art) & RCR25043104\\
        Yellow Ochre & Cranfield (Jackson's art) & RCR25091737\\
        Burnt Sienna & Cranfield (Jackson's art) & RCR50032371\\
        Raw Umber & Cranfield (Jackson's art) & RCR25032211\\
        White & Cranfield (Jackson's art) & RCR25083391\\
        SS Black & Culture Hustle (Stuart Semple) & BLACK 1.0 Pigment- 50g\\
        SS Red & Culture Hustle (Stuart Semple) & THE WORLD'S PINKEST PINK - 50g\\
        Hemoglobin & Sigma-Aldrich (Merck) & H2625-100G\\
        Nigrosin & Sigma-Aldrich (Merck) & 211680-100G\\
        IR-895 & Sigma-Aldrich (Merck) & 392375\\
        IR-1048 & Sigma-Aldrich (Merck) & 405175-500MG\\
        IR-1061 & Sigma-Aldrich (Merck) & 405124-250MG\\
        Spectrasense-765 & Sun Chemical Colors \& Effects GmbH & Spectrasense™ IR 765\\
        \hline
        \vspace{1pt}
    \end{tabular}
        
    \caption{Dyes that were investigated for their optical absorption and scattering spectra. The column "Dye name" indicates the name which is used in this work to refer to the specific dye and the "Product name" indicates the name that is used by the manufacturer/vendor.}
    \label{tab:dyes}
\end{table}

\begin{table}[H]
	\begin{center}
        \resizebox{\textwidth}{!}{
		\begin{tabular}{l c c c c c c c c c c c c}
			  & 1 & 2 & 3 & 4 & 5 & 6 & 7 & 8 & 9 & 10A & 10B & 10C\\
			\hline
			sO$_2$ & 0\% & 50\% & 100\% & 100\% & 50\% & 0\% & 100\% & 50\% & 0\% & 100\% & 0\% & 70\%\\
			Bvf & 2.5\% & 2.5\% & 2.5\% & 4\% & 4\% & 4\% & 1\% & 1\% & 1\% & 0.5\% & 5\% & 3\% \\
                \multirow{3}{0.5em}{sO$_2$ vessel} & & & & & & & & & & & &\\
                & 50\% & 30\% & 0\% & 70\% & 100\% & 100\% & / & / & / & / & / & /\\
                & & & & & & & & & & & & \\
			\hline
		\end{tabular}
        }
	\end{center}
    \caption{Oxygen saturation (sO$_2$) and blood volume fraction (Bvf) levels for the background material of each forearm phantom. Bvf refers to the percentage of the total volume that is made up of the dye mixture. While all the others have as uniform as possible properties, the 10th forearm was separated into three parts, with different properties, named 10A, 10B, and 10C. Six of the forearms have a superficial vessel, the oxygen saturation level of which is also reported.}
\end{table}

\end{spacing}